\documentclass[useAMS,usenatbib]{mnras}

\usepackage{myaasmacros}
\usepackage{graphicx}
\usepackage{ulem}
\usepackage[table]{xcolor}
\usepackage{amsmath}
\usepackage{caption}

\def\ltsima{$\; \buildrel < \over \sim \;$}
\def\simlt{\lower.5ex\hbox{\ltsima}}
\def\gtsima{$\; \buildrel > \over \sim \;$}
\def\simgt{\lower.5ex\hbox{\gtsima}}
\DeclareFontFamily{U}{mathx}{\hyphenchar\font45}
\DeclareFontShape{U}{mathx}{m}{n}{<-> mathx10}{}
\DeclareSymbolFont{mathx}{U}{mathx}{m}{n}
\DeclareMathAccent{\widebar}{0}{mathx}{"73}

\def\siglos{\sigma_{\text{LOS}}}

\def\betastar{\beta^*}
\def\dd{\text{d}}

\def\GravSphere{{\sc GravSphere}}
\def\GravSpherebf{{\bfseries\scshape GravSphere}}

\def\coreNFW{{\sc coreNFW}}

\def\EMCEE{{\sc emcee}}
\def\Barolo{{\sc $^{3\rm D}$Barolo}}

\def\betastar{\tilde{\beta}}
\def\vsone{v_{s1}}
\def\vstwo{v_{s2}}
\def\vlosfour{\langle v_{\rm LOS}^4 \rangle}

\usepackage{array}
\newcolumntype{L}[1]{>{\raggedright\let\newline\\\arraybackslash\hspace{0pt}}m{#1}}
\newcolumntype{C}[1]{>{\centering\let\newline\\\arraybackslash\hspace{0pt}}m{#1}}
\newcolumntype{R}[1]{>{\raggedleft\let\newline\\\arraybackslash\hspace{0pt}}m{#1}}

\title[Dark matter heats up in dwarf galaxies]{Dark matter heats up in dwarf galaxies}

\author[Read]{J. I. Read$^1$\thanks{E-mail: justin.inglis.read@gmail.com}, M. G. Walker$^2$, P. Steger$^3$\\
  $^1$Department of Physics, University of Surrey, Guildford, GU2 7XH, UK\\
  $^2$McWilliams Center for Cosmology, Department of Physics, Carnegie Mellon University, 5000 Forbes Ave., \\
  Pittsburgh, PA 15213, United States\\
  $^3$Institute for Astronomy, Department of Physics, ETH Z\"urich, Wolfgang-Pauli-Strasse 27, CH-8093 Z\"urich, Switzerland\\
}
 
\begin{document}

\maketitle

\begin{abstract}
Gravitational potential fluctuations driven by bursty star formation can kinematically `heat up' dark matter at the centres of dwarf galaxies. A key prediction of such models is that, at a fixed dark matter halo mass, dwarfs with a higher stellar mass will have a lower central dark matter density. We use stellar kinematics and HI gas rotation curves to infer the inner dark matter densities of eight dwarf spheroidal and eight dwarf irregular galaxies with a wide range of star formation histories. For all galaxies, we estimate the dark matter density at a common radius of 150\,pc, $\rho_{\rm DM}(150\,\mathrm{pc})$. We find that our sample of dwarfs falls into two distinct classes. Those that stopped forming stars over $6$\,Gyrs ago favour central densities $\rho_{\rm DM}(150\,\mathrm{pc})>10^8$\,M$_{\odot}$\,kpc$^{-3}$, consistent with cold dark matter cusps, while those with more extended star formation favour $\rho_{\rm DM}(150\,\mathrm{pc})<10^8$\,M$_{\odot}$\,kpc$^{-3}$, consistent with shallower dark matter cores. Using abundance matching to infer pre-infall halo masses, $M_{200}$, we show that this dichotomy is in excellent agreement with models in which dark matter is heated up by bursty star formation. In particular, we find that $\rho_{\rm DM}(150\,\mathrm{pc})$ steadily decreases with increasing stellar mass-to-halo mass ratio, $M_*/M_{200}$. Our results suggest that, to leading order, dark matter is a cold, collisionless, fluid that can be kinematically `heated up' and moved around.
\end{abstract}


\section{Introduction}\label{sec:intro}

The standard $\Lambda$ Cold Dark Matter ($\Lambda$CDM) cosmological model gives a remarkable description of the growth of structure in the Universe on large scales \citep[e.g.][]{2006Natur.440.1137S,2006ApJ...648L.109C,2013AJ....145...10D,2014A&A...571A..16P,2016JCAP...08..012B,2016MNRAS.456.2301W}. Yet, on smaller scales inside the dark matter halos of galaxies, there have been long-standing tensions \citep[e.g.][]{2017ARA&A..55..343B}. The oldest of these is the `cusp-core' problem. Pure dark matter (DM) structure formation simulations in $\Lambda$CDM predict a universal DM halo profile that has a dense `cusp' at the centre, with inner density $\rho_{\rm DM} \propto r^{-1}$ \citep{1991ApJ...378..496D,1996ApJ...462..563N}. By contrast, observations of gas rich dwarf galaxy rotation curves have long favoured DM `cores', with $\rho_{\rm DM} \sim {\rm constant}$ \citep{1994ApJ...427L...1F,1994Natur.370..629M,2010AdAst2010E...5D,2017MNRAS.467.2019R}.

The cusp-core problem has generated substantial interest over the past two decades because it may point to physics beyond the collisionless `Cold Dark Matter' (CDM) typically assumed to date. \citet{2000PhRvL..84.3760S} were the first to suggest that `Self Interacting Dark Matter' (SIDM) -- that invokes a new force acting purely in the dark sector -- could transform a dense cusp to a core through energy transfer between the DM particles \citep[e.g.][]{2013MNRAS.430...81R,2015MNRAS.453...29E,2016PhRvL.116d1302K,2017MNRAS.470.1542S,2017MNRAS.472.2945R}. Warm Dark Matter (WDM) has also been proposed as a solution to the cusp-core problem (e.g. \citealt{2000PhRvD..62f3511H,2001ApJ...556...93B,2001ApJ...559..516A,2014MNRAS.439..300L,2017MNRAS.470.1542S}, but see \citealt{2001ApJ...561...35D,2012MNRAS.424.1105M} and \citealt{2013MNRAS.430.2346S}). Other solutions include `fuzzy DM' \citep{2000PhRvL..85.1158H,2017PhRvD..95d3541H}, `fluid' DM \citep{2000ApJ...534L.127P} and `wave-like' DM \citep{2014NatPh..10..496S}.

However, there is a more prosaic explanation for the cusp-core problem. If gas is slowly accreted onto a dwarf galaxy and then suddenly removed (for example by stellar winds or supernovae feedback) this causes the DM halo to expand, irreversibly lowering its central density\footnote{Note that there is an alternative mechanism by which stars and gas can alter the inner DM density profile. \citet{2001ApJ...560..636E} were the first to suggest that dense gas clumps could impart angular momentum to the inner DM density profile by dynamical friction, causing a cusp to flatten to a core (see also \citealt{2009ApJ...698.2093D}, \citealt{2010ApJ...725.1707G} and \citealt{2011MNRAS.416.1118C} for more recent work on this). Such a mechanism still requires stellar feedback to then destroy these dense gas clumps. Otherwise, the inner stellar density that results would be too high to be consistent with observations \citep[e.g.][]{2015MNRAS.446.1820N}. The predictions from this class of model can be rather degenerate with `DM heating' due to potential fluctuations \citep{2016Ap&SS.361..162D} and it may well be that both act in tandem in dwarf galaxies. This remains an area of active research.} \citep{1996MNRAS.283L..72N}. \citet{2002MNRAS.333..299G} showed that, for reasonable gas fractions and collapse factors, the overall effect of this `DM heating' is small. However, if the effect repeats over several cycles of star formation, it accumulates, leading eventually to complete DM core formation \citep{2005MNRAS.356..107R}. Indeed, recent numerical simulations of dwarf galaxies that resolve the impact of individual supernovae on the interstellar medium find that the gas mass within the projected half light radius of the stars, $R_{1/2}$, naturally rises and falls on a timescale comparable to the local dynamical time\footnote{Fluctuations in the central gas mass need not be very large to excite DM heating, so long as they are repeated enough times. \citet{2016MNRAS.459.2573R} find in their simulations that the dynamical mass within $R_{1/2}$ fluctuates by just $\sim 10$\%, yet this is sufficient transform a DM cusp to a core within $R_{1/2}$.}, transforming an initial DM cusp to a core (e.g. \citealt{2008Sci...319..174M,2012MNRAS.421.3464P,2015MNRAS.454.2092O,2016MNRAS.456.3542T,2016MNRAS.459.2573R}, and for a review see \citealt{2014Natur.506..171P}). Such simulations have already made several testable predictions. \citet{2013MNRAS.429.3068T} show that the gas flows that transform DM cusps to cores lead to a bursty star formation history, with a peak-to-trough ratio of $5-10$ and a duty cycle comparable to the local dynamical time. Furthermore, the stars are dynamically `heated' similarly to the DM, leading to a stellar velocity dispersion that approaches the local rotational velocity of the stars ($v/\sigma \sim 1$) inside $R_{1/2}$. Both of these predictions are supported by observations of dwarf galaxies \citep[e.g.][]{1998AJ....116.1227D,2002AJ....123..813D,2007ApJ...659..331Y,2014MNRAS.441.2717K,2012ApJ...750...33L,2017MNRAS.465.2420W,2017MNRAS.466...88S}. Further evidences for `DM heating' come from the observed age gradients in dwarfs \citep{2016ApJ...820..131E}.

While there is strong evidence that dwarf galaxies have bursty star formation histories, this is only circumstantial evidence for DM heating. The real `smoking gun' for DM cusp-core transformations lies in another key prediction from recent numerical models: DM core formation requires several cycles of gas inflow and outflow \citep{2005MNRAS.356..107R,2012MNRAS.421.3464P}. Thus, at fixed halo mass, galaxies that have formed more stars (i.e. that have undergone more gas inflow-outflow cycles) will have a lower central DM density \citep{2012MNRAS.421.3464P,2012ApJ...759L..42P,2014MNRAS.437..415D,2015MNRAS.454.2092O,2015MNRAS.450.3920B,2016MNRAS.459.2573R,2017MNRAS.466L...1D,2018arXiv180607679B}. By contrast, solutions to the cusp-core problem that invoke exotic DM predict no relationship between the central DM densities of dwarfs and their star formation histories\footnote{Most exotic DM models designed to solve the cusp-core problem predict that {\it all} dwarfs should have a central DM core. However, there can be exceptions to this. In SIDM models with a high self-interaction cross section, for example, dark matter halos undergo `core collapse', increasing their central density at late times \citep[e.g.][]{2012MNRAS.423.3740V}. However, while this will lead to some stochasticity in the central DM density of dwarfs, it will not lead to any relationship between their central DM densities and their star formation histories.}.

Whether or not a dwarf will form a DM core depends primarily on the number and amplitude of gas inflow-outflow cycles, and on the amount of DM that needs to be evacuated from the centre of the dwarf to form the core. This can be posed in the form of an energy argument, whereby the total energy available to move gas around depends on the total stellar mass formed, $M_*$, while the energy required to unbind the DM cusp depends on the DM halo mass, $M_{200}$ \citep{2012ApJ...759L..42P}. Thus, whether or not a DM core will form in a given dwarf galaxy depends primarily on its stellar mass to halo mass {\it ratio}, $M_*/M_{200}$ \citep{2014MNRAS.437..415D}. However, since $M_{200}$ is challenging to extrapolate from the data, in this paper we consider also a proxy for the ratio $M_*/M_{200}$: the star formation `truncation time', $t_{\rm trunc}$. We define this to be the time when the dwarf's star formation rate (SFR) fell by a factor of two from its peak value\footnote{This is similar to the concept of `fast' and `slow' dwarfs introduced by \citet{2015ApJ...811L..18G} and explored in more detail by \citet{2018arXiv180607679B}. However, our definition here is more readily applied to our sample of both dSphs and dIrrs (see also \citet{2018arXiv180707093R} for a discussion on this point).}. This can be used as a proxy for $M_*/M_{200}$ so long as the SFR is approximately constant\footnote{If dwarfs have significant gaps in their star formation histories, then this correspondence between $t_{\rm trunc}$ and $M_*/M_{200}$ can break \citep[e.g.][]{2019MNRAS.482.1176W}. For this reason, in this paper we will look for anti-correlations between the central DM density of dwarfs and $t_{\rm trunc}$ (that is easier to measure) and $M_*/M_{200}$ (that is more fundamental, but harder to estimate).} (as is the case for the sample of dwarfs that we consider in this paper; see \citealt{2018arXiv180707093R} and \S\ref{sec:data}). In this case, dwarfs with $t_{\rm trunc} \rightarrow 0$\,Gyrs have $M_*/M_{200} \rightarrow 0$, while those with $t_{\rm trunc} \rightarrow 13.8$\,Gyrs (i.e. the age of the Universe) have formed stars for as long as possible and have, therefore, maximised both $M_*/M_{200}$ and their ability to produce a DM core. Unlike $M_{200}$, however, $t_{\rm trunc}$ has the advantage that it is readily estimated from the dwarf's star formation history (SFH; see \S\ref{sec:data}).

In this paper, we set out to test the above key prediction of DM heating models, that dwarfs with `extended star formation' (i.e. $t_{\rm trunc} \rightarrow 13.8$\,Gyrs and maximal $M_*/M_{200}$) have DM cores, while those with `truncated star formation' (i.e. $t_{\rm trunc} \rightarrow 0$\,Gyrs and minimal $M_*/M_{200}$) have DM cusps. To achieve this, we estimate the central DM density, $M_*$, $t_{\rm trunc}$ and $M_{200}$ for a sample of nearby dwarf galaxies with a wide range of star formation histories (SFHs). Our sample includes gas-poor dwarf spheroidal galaxies (dSphs) whose star formation ceased shortly after the beginning of the Universe, dSphs with extended star formation that shut down only very recently, and gas rich dwarf irregular galaxies (dIrrs) that are still forming stars today. This requires us to accurately infer the DM distribution in both gas rich and gas poor galaxies. For the former, we use HI rotation curves as in \citet{2017MNRAS.467.2019R}; for the latter, we use line of sight stellar kinematics. However, with only line of sight velocities, there is a well-known degeneracy between the radial density profile (that we would like to measure) and the velocity anisotropy of the dwarf (see \S\ref{sec:gravsphere} and \citealt{1982MNRAS.200..361B,1990AJ.....99.1548M,2013NewAR..57...52B,2017MNRAS.471.4541R}). In \citet{2017MNRAS.471.4541R} and \citet{Read:2018pft}, we introduced a new mass modelling tool -- \GravSphere\ -- that breaks this degeneracy by using `Virial Shape Parameters' (VSPs). We used a large suite of mock data to demonstrate that with $\sim 500$ radial velocities, \GravSphere\ is able to correctly infer the dark matter density profile over the radial range $0.5 < r/R_{1/2} < 2$, within its 95\% confidence intervals. Here, we use \GravSphere\ to infer the inner DM density of eight Milky Way dSphs that have radial velocities for $\simgt 190$ member stars. We emphasise that, while with of order 500 radial velocities, \GravSphere\ is not able to obtain a robust inference of the inner {\it slope} of the DM density profile, it can constrain the {\it amplitude} of the inner DM density at $\sim 150$\,pc \citep{Read:2018pft}. As we shall show, this is sufficient to test DM heating models.

This paper is organised as follows. In \S\ref{sec:cuspcoretheory}, we briefly review the cusp-core problem in $\Lambda$CDM, and we explain why a robust estimate of the amplitude of the DM density at 150\,pc is sufficient for testing DM heating models. In \S\ref{sec:method}, we describe our method for measuring the DM density profile from stellar kinematics (\GravSphere; \S\ref{sec:gravsphere}) and HI rotation curves (\S\ref{sec:HIrot}). In \S\ref{sec:data}, we describe our data compilation for our sample of dIrrs and dSphs, including their SFHs and estimates of $M_{200}$ taken from the literature. In \S\ref{sec:results}, we present our key results. In \S\ref{sec:discussion}, we compare our measurements with previous work in the literature. We discuss the robustness of our results and their implications for `DM heating' and the nature of DM. Finally, in \S\ref{sec:conclusions} we present our conclusions.

\section{The cusp-core problem in $\Lambda$CDM}\label{sec:cuspcoretheory}

In this section, we briefly review the cusp-core problem in $\Lambda$CDM. This broadly follows a similar review presented in \citet{Read:2018pft}; however, we reproduce this here in order to introduce some key equations that we will need later on, and for this paper to be self-contained. 

Pure DM structure formation simulations in $\Lambda$CDM predict DM halos that have a `Navarro, Frenk \& White' (NFW) density profile \citep{1996ApJ...462..563N}: 

\begin{equation} 
\rho_{\rm NFW}(r) = \rho_0 \left(\frac{r}{r_s}\right)^{-1}\left(1 + \frac{r}{r_s}\right)^{-2}
\label{eqn:rhoNFW}
\end{equation}
where the central density $\rho_0$ and scale length $r_s$ are given by: 
\begin{equation} 
\rho_0 = \rho_{\rm crit} \Delta c_{200}^3 g_c / 3 \,\,\,\, ; \,\,\,\, r_s = r_{200} / c_{200}
\end{equation}
\begin{equation}
g_c = \frac{1}{{\rm log}\left(1+c_{200}\right)-\frac{c_{200}}{1+c_{200}}}
\end{equation}
and:
\begin{equation} 
r_{200} = \left[\frac{3}{4} M_{200} \frac{1}{\pi \Delta \rho_{\rm crit}}\right]^{1/3}
\label{eqn:r200}
\end{equation} 
where $c_{200}$ is the concentration parameter; $\Delta = 200$ is the over-density parameter; $\rho_{\rm crit} = 136.05$\,M$_\odot$\,kpc$^{-3}$ is the critical density of the Universe at redshift $z=0$; $r_{200}$ is the virial radius; and $M_{200}$ is the virial mass.

The mass and concentration of halos in $\Lambda$CDM are correlated \citep[e.g.][]{2014MNRAS.441.3359D}: 

\begin{equation}
\log_{10}(c_{200}) = 0.905 - 0.101 \log_{10}(M_{200} h - 12)
\label{eqn:M200c200}
\end{equation}
with scatter $\Delta \log_{10}(c_{200}) = 0.1$, where $h \sim 0.7$ is the Hubble parameter. 

\begin{figure}
\begin{center}
\includegraphics[width=0.45\textwidth]{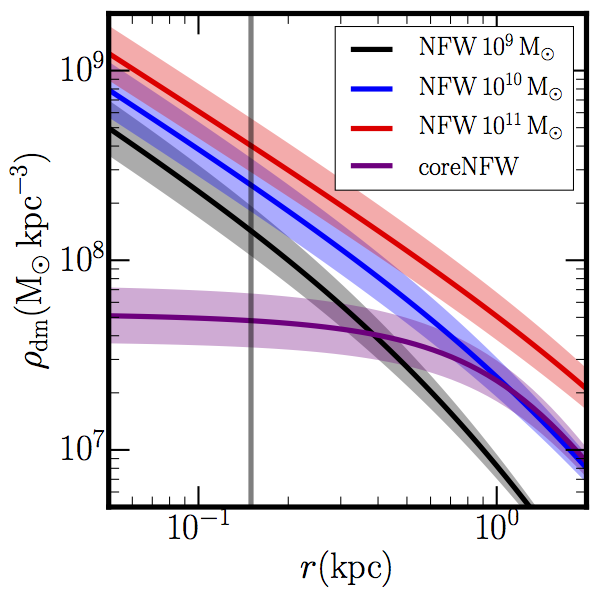}
\caption{Dark matter density profiles in $\Lambda$CDM. The black, blue and red lines show results from fits to pure DM structure formation simulations (i.e. NFW profiles; equation \ref{eqn:rhoNFW}) for three halo masses: $M_{200} = 10^9\,{\rm M}_\odot, 10^{10}\,{\rm M}_\odot$ and $10^{11}\,{\rm M}_\odot$, as marked in the legend. The purple line shows a fit to a model in which star formation `heats up' the DM halo, lowering its central density (i.e. a \coreNFW\ profile; equation \ref{eqn:rhocNFW}). The median lines assume that halos lie on the $M_{200}-c_{200}$ relation (equation \ref{eqn:M200c200}), while the shaded regions show the $1\sigma$ scatter in this relation. Notice that the central densities of the cusped (black, blue and red) and cored (purple) models is very different. A single measurement of the density at $150$\,pc (vertical grey line) is sufficient to differentiate the models, independently of the halo mass.}
\label{fig:NFW_similar}
\end{center}
\end{figure}

Recent simulations, that have sufficient spatial resolution to capture the dense multiphase interstellar medium ($\Delta x \simlt 100$\,pc), and that include the effects of gas cooling, star formation and feedback, find that DM cusps are transformed to cores in the centres of dwarf galaxies \citep[e.g.][]{2008Sci...319..174M,2012MNRAS.421.3464P,2013MNRAS.429.3068T,2014Natur.506..171P,2014MNRAS.437..415D,2015MNRAS.454.2092O,2016MNRAS.459.2573R}. \citet{2016MNRAS.459.2573R} introduced a fitting function to parameterise this cusp-core transformation, the `\coreNFW' profile. This has a cumulative mass profile given by:

\begin{equation}
M_{\rm cNFW}(<r) = M_{\rm NFW}(<r) f^n
\label{eqn:coreNFW}
\end{equation}
where $M_{\rm NFW}(<r)$ is the NFW cumulative mass profile: 

\begin{equation} 
M_{\rm NFW}(r) = M_{200} g_c \left[\ln\left(1+\frac{r}{r_s}\right) - \frac{r}{r_s}\left(1 + \frac{r}{r_s}\right)^{-1}\right]
\label{eqn:MNFW}
\end{equation}
and $f^n$ generates a shallower density profile at radii $r \simlt r_c$:  

\begin{equation} 
f^n = \left[\tanh\left(\frac{r}{r_c}\right)\right]^n
\end{equation}
The density profile of the \coreNFW\ model is given by:

\begin{equation} 
\rho_{\rm cNFW}(r) = f^n \rho_{\rm NFW} + \frac{n f^{n-1} (1-f^2)}{4\pi r^2 r_c} M_{\rm NFW}
\label{eqn:rhocNFW}
\end{equation} 
(The other main fitting function proposed in the literature to date -- the \citealt{2014MNRAS.441.2986D} profile -- produces similar results when applied to both simulated and real data; \citealt{2017MNRAS.470.1542S,2017A&A...605A..55A}.)

In Figure \ref{fig:NFW_similar}, we show fits to the DM density profiles of halos extracted from pure DM cosmological simulations in $\Lambda$CDM, with virial masses over range: $10^9 < M_{200}/{\rm M}_\odot < 10^{11}$, corresponding to dwarf galaxies. The median lines assume that halos lie on the $M_{200}-c_{200}$ relation (equation \ref{eqn:M200c200}), while the shaded regions show the $1\sigma$ scatter in this relation. The purple line shows a maximally cored DM halo (equation \ref{eqn:rhocNFW}) with $M_{200} = 10^{10}$\,M$_\odot$, $n=1$ and $R_{1/2} = 0.015\, r_{200} = 0.7$\,kpc \citep{2013ApJ...764L..31K}, corresponding to $r_c = 1.2$\,kpc. This cored model gives a good match to simulations in which DM cusps are transformed to cores by gas flows \citep{2016MNRAS.459.2573R}, but also to models in which cores form due to self-interactions between DM particles \citep{2017MNRAS.470.1542S,Read:2018pft}. The key difference between these two models, as highlighted in \S\ref{sec:intro}, is that the former predicts an anti-correlation between the DM core size and the total amount of star formation in a dwarf galaxy (i.e. the longer the star formation continues, the more times gas cycles in and out of the centre of the dwarf and the more DM heating occurs), while the latter predicts no such anti-correlation. This is the key difference that we set out to test in this paper.

The striking thing to note from Figure \ref{fig:NFW_similar} is just how different the central densities of the cored and cusped models are, independently of halo mass $M_{200}$. While a measurement of the {\it slope} of the density profile is ideal for differentiating models, we can actually differentiate amongst these cored and cusped models with a single measurement of the density at small radii. In this paper, we choose this `small radius' to be $r_{\rm S} = 150$\,pc (vertical grey line). This represents a compromise between picking $r_{\rm S}$ small enough to differentiate between interesting models, but not so small that the uncertainties on $\rho_{\rm DM}(r_{\rm S})$ are prohibitively large. In Appendix \ref{app:scale_test}, we show that our results are not sensitive to this choice of $r_{\rm S}$.

The inner logarithmic slope of the density profile, $\gamma_{\rm DM}(r_S) 
\equiv d \ln \rho_{\rm DM} / d\ln r(r_S)$, or the asymptotic slope, $\gamma_{\rm DM}(r \rightarrow 0)$, have traditionally be used to differentiate cored and cusped models \citep[e.g.][]{2013MNRAS.433.2314H}. However, as can be seen in Figure \ref{fig:NFW_similar}, we can obtain useful cosmological information also from the amplitude of the DM density profile at $r_S = 150\,{\rm pc}$. In \citet{Read:2018pft}, we used mock data for a Draco-like dwarf to show that, with $\sim 500$ stars with radial velocities, \GravSphere's inference of $\gamma_{\rm DM}(150\,{\rm pc})$ depended on our choice of priors on $\gamma_{\rm DM}$. By contrast, \GravSphere's inference of $\rho_{\rm DM}(150\,{\rm pc})$ was not sensitive to these priors. For this reason, we focus in this paper primarily on $\rho_{\rm DM}(150\,{\rm pc})$. For completeness, we show results for $\gamma_{\rm DM}(150\,{\rm pc})$ in Appendix \ref{app:gammaDM}.

\section{Method}\label{sec:method}

\subsection{Modelling the stellar kinematics: \GravSphere}\label{sec:gravsphere}

\GravSphere\ is described and tested in detail in \citet{2017MNRAS.471.4541R} and \cite{Read:2018pft}. It solves the projected spherical Jeans equation \citep{1922MNRAS..82..122J,1982MNRAS.200..361B}:

\begin{equation}
 \siglos^2(R) = \frac{2}{\Sigma(R)}\int_R^\infty \left(1\!-\!\beta\frac{R^2}{r^2}\right)
    \nu\sigma_r^2\,\frac{r\,\dd r}{\sqrt{r^2\!-\!R^2}} \ ,
    \label{eqn:LOS}
\end{equation}
where $\Sigma(R)$ denotes the tracer surface mass density at projected radius $R$; $\nu(r)$ is the spherically averaged tracer density; and $\beta(r)$ is the velocity anisotropy:

\begin{equation}
\beta = 1 - \frac{\sigma_t^2}{\sigma_r^2}
\label{eqn:beta}
\end{equation}
where $\sigma_t$ and $\sigma_r$ are the tangential and radial velocity dispersions, respectively, and $\sigma_r$ is given by \citep{1994MNRAS.270..271V,2005MNRAS.362...95M}:
\begin{equation}
\sigma_r^2(r) = \frac{1}{\nu(r) g(r)} \int_r^\infty \frac{GM(\tilde{r})\nu(\tilde{r})}{\tilde{r}^2} g(\tilde{r}) \dd \tilde{r}
\label{eqn:main}
\end{equation}
where:
\begin{equation}
g(r) = \exp\left(2\int \frac{\beta(r)}{r}\dd r\right)
\label{eqn:ffunc}
\end{equation}
and $M(r)$ is the cumulative mass of the dwarf galaxy (due to all stars, gas, DM etc.), that we would like to measure.

\GravSphere\ uses a non-parametric model for $M(r)$ that comprises a contribution from all visible matter and a contribution from DM that is described by a sequence of power laws defined on a set of radial bins. In this paper, these bins are defined at $[0.25,0.5,1,2,4]R_{1/2}$, where $R_{1/2}$ is the projected half light radius of the tracer stars. The tracer light profile is also non-parametric, using a series sum of Plummer spheres, as in \citet{2016MNRAS.459.3349R}. The velocity anisotropy is given by a form that makes $g(r)$ analytic: 

\begin{equation} 
\beta(r) = \beta_0 + \left(\beta_\infty-\beta_0\right)\frac{1}{1 + \left(\frac{r_0}{r}\right)^n}
\label{eqn:betagravsphere}
\end{equation}
where $\beta_0$ is the inner asymptotic anisotropy, $\beta_\infty$ is the outer asymptotic anisotropy, $r_0$ is a transition radius, and $n$ controls the sharpness of the transition.

We use a symmetrised $\betastar$ \citep{2006MNRAS.367..387R,2017MNRAS.471.4541R}:

\begin{equation} 
\betastar =  \frac{\sigma_r^2 - \sigma_t^2}{\sigma_r^2 + \sigma_t^2} = \frac{\beta}{2-\beta}
\label{eqn:betastar}
\end{equation} 
since this avoids infinities in $\beta$ for highly tangential orbits. We assume flat priors on $-1 < \betastar_{0,\infty} < 1$ such that we give equal weight to tangentially and radially anisotropic models.

By default, \GravSphere\ also fits for the two higher order `Virial Shape Parameters' (VSPs; \citealt{1990AJ.....99.1548M,2014MNRAS.441.1584R,2017MNRAS.471.4541R}):

\begin{eqnarray} 
\vsone & = & \frac{2}{5}\, \int_0^{\infty} G M\, (5-2\beta) \,\nu \sigma_r^2 \,r \,\dd r \\
\label{eqn:vs1}
& = & \int_0^{\infty} \Sigma \vlosfour\, R\, \dd R
\label{eqn:vs1data}
\end{eqnarray}
and:
\begin{eqnarray} 
\vstwo & = & \frac{4}{35} \,\int_0^{\infty} G M\, (7-6\beta)\, \nu \sigma_r^2 \,r^3 \,\dd r \\
\label{eqn:vs2}
& = & \int_0^{\infty} \Sigma \vlosfour\, R^3\, \dd R \ .
\label{eqn:vs2data}
\end{eqnarray}
These allow \GravSphere\ to break the $\rho-\beta$ degeneracy \citep{2017MNRAS.471.4541R}. We use the improved estimators for $\vsone$ and $\vstwo$ described in \cite{Read:2018pft}.

\GravSphere\ fits the above model to the surface density profile of tracer stars, $\Sigma_*(R)$, their line-of-sight projected velocity dispersion profile $\siglos(R)$ and their VSPs using the \EMCEE\ affine invariant Markov Chain Monte Carlo (MCMC) sampler from \citet{2013PASP..125..306F}. We assume uncorrelated Gaussian errors such that the Likelihood function is given by $\mathcal{L} = \exp(-\chi^2/2)$, where $\chi^2$ includes the contributions from the fits to $\Sigma_*, \siglos$ and the two VSPs. We use 1000 walkers, each generating 5000 models and we throw out the first half of these as a conservative `burn in' criteria. (See \citet{2017MNRAS.471.4541R} and \citet{Read:2018pft} for further details of our methodology and priors.)

\GravSphere\ has been extensively tested on mock data with realistic contamination and selection criteria, realistic departures from spherical symmetry and realistic disequilibrium due to tidal stripping \citep{2017MNRAS.471.4541R,Read:2018pft}. In all cases, \GravSphere\ was able to recover the key quantity of interest for this paper -- $\rho_{\rm DM}(150\,{\rm pc})$ -- within its 95\% confidence intervals.

\subsection{Fitting gaseous rotation curves}\label{sec:HIrot}

For the gas rich isolated dwarfs, we derive the rotation curves from HI datacubes using the \Barolo\ software, as in \citet{2017MNRAS.467.2019R} and \citet{2017MNRAS.466.4159I}. For the mass model, we decompose the circular speed curve into contributions from stars, gas and DM: 

\begin{equation}
v_c^2 = v_*^2 + v_{\rm gas}^2 + v_{\rm dm}^2
\end{equation}
where $v_*$ and $v_{\rm gas}$ are the contributions from stars and gas, respectively, and $v_{\rm dm}$ is the DM contribution. 

We assume that both the stars and gas are exponential discs: 

\begin{equation} 
v_{*/{\rm gas}}^2 = \frac{2 G M_{*/{\rm gas}}}{R_{*/{\rm gas}}} y^2 \left[I_0(y) K_0(y) - I_1(y) K_1(y)\right]
\label{eqn:vcstargas}
\end{equation}
where $M_{*/{\rm gas}}$ is the mass of the star/gas disc, respectively; $R_{*/{\rm gas}}$ is the exponential scale length; $y = R/R_{*/{\rm gas}}$ is a dimensionless radius parameter; and $I_0, I_1, K_0$ and $K_1$ are Bessel functions \citep{1987gady.book.....B}. As in \citet{2017MNRAS.467.2019R}, we fix the values of $R_*$, $R_{\rm gas}$ and $M_{\rm gas}$ to the median of their observed values in our model fits. All values used are reported in Table \ref{tab:data}.

To ensure consistency between the stellar kinematic and gas rich models that we present here, for the DM mass distribution ($v_{\rm dm}^2 = G M_{\rm dm} / r$), we use the freeform mass model from \citet{2017MNRAS.471.4541R}, described in \S\ref{sec:gravsphere}, above. This differs from the analysis in \citet{2016MNRAS.462.3628R} and \citet{2017MNRAS.467.2019R} where we used instead the `coreNFW' profile from \citet{2016MNRAS.459.2573R}. In tests, we verified that this choice does not affect our results. (Using the coreNFW distribution instead, and allowing the core-size parameter, $r_c$, to freely vary, leads to density profiles consistent with our free-form models, but with smaller uncertainties corresponding to the reduced freedom in the mass model.)

\section{The Data}\label{sec:data}

Our data sample comprises nearby dwarf galaxies that -- based on mock data tests -- have sufficiently good data to estimate $\rho_{\rm DM}(150\,{\rm pc})$ reliably, and that have had their data analysed in a homogeneous manner. These are the eight Milky Way `classical' dSphs \citep[e.g.][]{2012AJ....144....4M}, and eight isolated gas rich dIrr galaxies taken from \citet{2017MNRAS.467.2019R}. 

\subsection{The dwarf irregulars}
For the isolated dIrrs, we measure their DM density profile from their HI gas rotation curves, as described in \citet{2017MNRAS.467.2019R} and \S\ref{sec:HIrot}. The rotation curves for these galaxies were extracted from the HI datacubes using \Barolo, as described in detail in \citet{2017MNRAS.467.2019R} and \citet{2017MNRAS.466.4159I}. As in \citet{2017MNRAS.467.2019R}, our isolated dwarf sample is chosen to have an inclination angle of $i > 40^\circ$ because \Barolo\ can become systematically biased for lower inclination angles than this. We also require a good measurement of the distance and photometric light profile \citep{2016MNRAS.462.3628R}. Two of the dwarfs, WLM and Aquarius, have star formation histories derived from deep colour magnitude diagrams \citep{2000ApJ...531..804D,2014ApJ...795...54C}; the remainder are known to be still forming stars today \citep{2012AJ....143...47Z}. Finally, of the 11 dwarfs in \citet{2017MNRAS.467.2019R} that meet the above criteria, we exclude NGC 6822 because it has a central stellar bar that complicates the analysis, and DDO 126 and UGC 8505 because their inner rotation curves are sufficiently uncertain that we are unable to obtain a good measurement of $\rho_{\rm DM}(150\,{\rm pc})$. The data for our sample of dIrrs is described and presented in detail in \citet{2017MNRAS.467.2019R} and \citet{2017MNRAS.466.4159I} and so we refer the reader to those publications for further details.

\subsection{The dwarf spheroidals}
Our sample of dSphs: Draco, UMi, Sculptor, Carina, Fornax, Sextans, Leo I and Leo II, each have $\simgt 190$ stars with radial velocities and well-measured photometric light profiles. The best-sampled systems have over 500 member velocities (Draco [504], Carina [767], Sculptor [1,351] and Fornax [2,573]). We mass-model these dSphs using the \GravSphere\ code (see \citealt{2017MNRAS.471.4541R,Read:2018pft} and \S\ref{sec:gravsphere}). With $\sim 500$ member velocities, \GravSphere\ can estimate $\rho_{\rm DM}(150\,{\rm pc})$ well enough to distinguish a $\Lambda$CDM cusp from a constant density core at 95\% confidence (see \S\ref{sec:cuspcoretheory} and \S\ref{sec:gravsphere}). \GravSphere\ gracefully degrades as the number of data points are reduced.

Since \GravSphere\ simultaneously fits both surface density and projected velocity dispersion profiles, for each dSph we require both photometric and kinematic data. For the photometric data, we use the Pan-STARRS DR1 catalog \citep{flewelling16} for the northern dwarfs Draco, Leo I, Leo II, Sextans and Ursa Minor. For the southern dwarfs Fornax and Scuptor we use data from the VLT/ATLAS DR1 catalog, as re-processed and calibrated by \citet{koposov14}. For the southern dwarf Carina, which is not included in either of the above catalogs, we use a catalog derived from observations with the Dark Energy Camera by \citet{mcmonigal14} and generously provided by those authors (N. McMonigal, private comm.). From each photometric catalog we initially select point-like sources\footnote{For the Pan-STARRS catalogs we select point sources as objects for which the difference between PSF and Kron magnitudes in the $r$ band is $r_{\rm PSF}-r_{\rm kron}<0.05$ (see \citealt{farrow14} for a discussion of Pan-STARRS star-galaxy separation). For the ATLAS catalogs we select objects classified as stars (star/galaxy classifier value of $-1$). For Carina we use objects classified as stars by \citet{mcmonigal14}.} within circular apertures of sufficient angular radius ($1.5^{\circ}$ for each of Draco, Fornax, Sculptor, Sextans and Ursa Minor; $1^{\circ}$ for Leo I and Leo II; $0.9^{\circ}$ for Carina) to enclose all plausibly-bound member stars. From these point sources we obtained samples of candidate red giant branch (RGB) stars within each dwarf galaxy by selecting only sources that are brighter than $i\leq 21$ mag and that deviate in colour-magnitude ($g-r$, $i$) space by less than $\epsilon$ magnitudes from an old (age=12 Gyr), metal-poor ([Fe/H]=-2.5) model isochrone \citep{dotter08} that we shift by the distance modulus corresponding to each galaxy's published distance \citep{2012AJ....144....4M}. The only exception is Carina, for which $i$-band data are not available and we use $g$ instead, keeping the same magnitude limit of $g\leq 22$. For this work, we adopt $\epsilon=\sqrt{0.04+\sigma^2_i+\sigma^2_{g-r}}$, where $\sigma_i$ and $\sigma_{g-r}$ are the photometric uncertainties in magnitude and colour, respectively.  

For the stellar-kinematic data, we use the published spectroscopic samples of \citet{2009AJ....137.3100W} for Carina, Fornax, Sculptor and Sextans, of \citet{mateo08} for Leo I, of \citet{spencer17} for Leo II, and of \citet{2015MNRAS.448.2717W} for Draco. For Ursa Minor, we use spectroscopic data that were acquired, processed and analysed in the same way as that of Draco (Spencer et al., in preparation). In addition to line-of-sight velocities, these data sets contain information about the chemical composition of individual stars, in the form either of a magnesium index \citep{2009AJ....137.3100W} or a direct estimate of [Fe/H] \citep{2015MNRAS.448.2717W}; the only exception is Leo I, for which only velocities are available. In order to separate dwarf galaxy members from contamination from the Galactic foreground, we fit an initial, chemodynamical mixture model that is similar to the one described in detail by \citet{caldwell17}; the only difference is that here we assume any velocity and/or metallicity gradients are negligible.  After fitting these simple models, we evaluate for every individual star a probability of dwarf galaxy membership, $P_{\rm mem}$, according to Equation 7 of \citet[][ for RGB candidates lacking spectroscopic measurements, we evaluate membership probability based only on projected distance from the dwarf galaxy centre]{caldwell17}. We then construct empirical surface density and projected velocity dispersion profiles by dividing the photometric and spectroscopic data sets, respectively, into annular bins that each contain equal numbers (weighted by membership probability) of member stars. We confirm that our results are qualitatively unchanged for alternative profiles that use different numbers of bins and/or membership probabilities obtained from more sophisticated initial models (e.g., ones that explicitly allow for radially varying velocity dispersion).  

\subsection{Star formation histories}

For the star formation histories (SFHs), where possible we use literature determinations derived from deep resolved colour magnitude diagrams (Draco, \citealt{2001AJ....122.2524A}; Sculptor, \citealt{2012A&A...539A.103D}; Carina, \citealt{2014A&A...572A..10D}; Fornax, \citealt{2012A&A...544A..73D}; Sextans, \citealt{2009ApJ...703..692L}; UMi, \citealt{2002AJ....123.3199C}; Leo I, \citealt{2002MNRAS.332...91D}, Leo II, \citealt{2002MNRAS.332...91D}, WLM, \citealt{2000ApJ...531..804D}; and Aquarius, \citealt{2014ApJ...795...54C}). For the remainder of our sample of dIrrs -- that are all still star forming today -- we use the SFH measured from their integrated light by \citet{2012AJ....143...47Z}.

\subsection{Dark matter halo masses}

We obtain $M_{200}$ for our sample of dIrrs by using an extrapolation from their HI rotation curves \citep{2017MNRAS.467.2019R}. For our sample of dSphs, we use estimates from a novel form of abundance matching that corrects for satellite quenching on infall \citep{2018arXiv180707093R}. As discussed in \citet{2017MNRAS.467.2019R} and \citet{2018arXiv180707093R} these abundance matching estimates of $M_{200}$ agree remarkably well with dynamical estimates from HI rotation curves or stellar kinematics. In \S\ref{sec:results}, we show that our results are not sensitive to even rather large systematic errors in our estimates of $M_{200}$.

Our full data sample, including half light radii, stellar masses, HI masses, stellar kinematic sample size and data references are given in Table \ref{tab:data}. There, we also report $t_{\rm trunc}$ for each dwarf (see \S\ref{sec:intro}), $M_{200}$ (see above) and our estimates of $\rho_{\rm DM}(150\,{\rm pc})$ (see \S\ref{sec:results}) and $\gamma_{\rm DM}(150\,{\rm pc})$ (see Appendix \ref{app:gammaDM}).

\begin{table*}
\begin{center}
\resizebox{\textwidth}{!}{
\begin{tabular}{L{1.5cm} l | c c c c c c c c c c | l}
\hline
\hline
{\bf Galaxy} \vspace{1mm} & {\bf Type} & $\mathbf{D}$ & $\mathbf{M_*}$ & $\mathbf{M_{\rm gas}}$ & $\mathbf{R_{1/2}}$ & $\mathbf{R_{\rm gas}}$ & $\mathbf{M_{200}}$ & {\bf Sample} & $\mathbf{t_{\rm trunc}}$ & $\rho_{\rm DM}(150\,{\rm pc})$ & $\gamma_{\rm DM}(150\,{\rm pc})$ & {\bf Refs.} \\
& & (kpc) & $(10^6\,{\rm M}_\odot)$ & $(10^6\,{\rm M}_\odot)$ & (kpc) & (kpc) & $(10^{9} {\rm M}_\odot)$ & {\bf size} & (Gyrs) & ($10^8 {\rm M}_\odot\,{\rm kpc}^{-3}$) & & \\
\hline
UMi & dSph & $76\pm 3$ & $0.29$ & -- & $0.181 \pm 0.027 \,[0.306]$ & -- & $2.8 \pm 1.1$ & 430 & 12.4 & $1.53_{-0.32}^{+0.35}$ & $-0.71_{-0.29}^{+0.28}$ & 3,5 \\ [2ex]
Draco & dSph & $76\pm 6$ & $0.29$ & -- & $0.221 \pm 0.019 \,[0.198]$ & -- & $1.8 \pm 0.7$ & 504 & 11.7 & $2.36_{-0.29}^{+0.29}$ & $-0.95_{-0.25}^{+0.25}$ & 3,4 \\ [2ex]
Sculptor & dSph & $86\pm 6$ & $2.3$ & -- & $0.283\pm 0.045 \,[0.248]$ & -- & $5.7 \pm 2.3$ & 1,351 & 11.8 & $1.49_{-0.23}^{+0.28}$ & $-0.83_{-0.25}^{+0.3}$ & 3,6 \\ [2ex]
Sextans & dSph & $86\pm 4$ & $0.44$ & -- & $0.695 \pm 0.044 \,[0.352]$ & -- & $2.0 \pm 0.8$ & 417 & 10.6 & $1.28_{-0.29}^{+0.34}$ & $-0.95_{-0.41}^{+0.36}$ & 3,7 \\ [2ex]
Leo I & dSph & $254\pm15$ & $5.5$ & -- & $0.251\pm 0.027 \,[0.298]$ & -- &  $5.6 \pm 2.2$ & 328 & 3.1 & $1.77_{-0.34}^{+0.33}$ & $-1.15_{-0.37}^{+0.33}$ & 3,8 \\ [2ex]
Leo II & dSph & $233\pm14$ & $0.74$ & -- & $0.176\pm 0.042 \,[0.194]$ & -- & $1.6 \pm 0.7$ & 186 & 6.3 & $1.84_{-0.16}^{+0.17}$ & $-1.5_{-0.31}^{+0.35}$ & 3,8 \\ [2ex]
Carina & dSph & $105\pm6$ & $0.38$ & -- & $0.250\pm 0.039 \,[0.242]$ & -- & $0.8 \pm 0.30$ & 767 & 2.8 & $1.16_{-0.22}^{+0.20}$ & $-1.23_{-0.35}^{+0.39}$ & 3,9 \\ [2ex]
Fornax & dSph & $138\pm 8$ & $43$ & -- & $0.710\pm 0.077 \,[0.670]$ & -- & $21.9 \pm 7.4$ & 2,573 & 1.75 & $0.79_{-0.19}^{+0.27}$ & $-0.30_{-0.28}^{+0.21}$ & 3,10 \\ [2ex]
\hline
WLM & dIrr & $985\pm 33$ & $16.2 \pm 4$ & $79$ & $1.26$ & $1.04$ & $8.3_{-2}^{+2}$ & -- & 0 & $0.52_{-0.09}^{+0.09}$ & $-0.37_{-0.16}^{+0.19}$ & 1,2 \\ [2ex]
DDO 52 & dIrr & 10,300 & $52.7 \pm 13$ & $371$ & 1.58 & 2.49 & $12_{-2.7}^{+2.9}$ & -- & 0 & $0.38_{-0.10}^{+0.17}$ & $-0.18_{-0.24}^{+0.13}$ & 1 \\ [2ex]
DDO 87 & dIrr & 7,400 & $33 \pm 8$ & $310$ & 1.9 & 1.51 & $11.3_{-2.5}^{+2.7}$ & -- & 0 & $0.31_{-0.09}^{+0.18}$ & $-0.22_{-0.24}^{+0.15}$ & 1 \\ [2ex]
DDO 154 & dIrr & 3,700 & $8.35 \pm 2$ & $309$ & 0.91 & 2.34 & $12.6_{-0.5}^{+0.5}$ & -- & 0 & $0.46_{-0.10}^{+0.13}$ & $-0.20_{-0.24}^{+0.15}$ & 1 \\ [2ex]
Aquarius & dIrr & 900 & $0.68 \pm 0.17$ & $3.3$ & 0.37 & 0.25 & $0.68_{-0.4}^{+1.3}$ & -- & 0 & $0.36_{-0.19}^{+0.22}$ & $-0.41_{-0.51}^{+0.31}$ & 1 \\ [2ex]
NGC 2366 & dIrr & 3,400 & $69.5 \pm 17.3$ & $1,730$ & 1.54 & 2.69 & $24_{-5.4}^{+4.9}$ & -- & 0 & $0.18_{-0.03}^{+0.05}$ & $-0.09_{-0.12}^{+0.07}$ & 1 \\ [2ex]
CVnIdwA & dIrr & 3,600 & $4.1 \pm 1$ & $64.2$ & 1.14 & 1.18 & $1.7_{-0.5}^{+1}$ & -- & 0 & $0.33_{-0.09}^{+0.12}$ & $-0.25_{0.27}^{+0.17}$ & 1 \\ [2ex]
DDO 168 & dIrr & 4,300 & $59 \pm 14.8$ & $458$ & 1.38 & 1.51 & $21_{-4.8}^{+5.2}$ & -- & 0 & $0.31_{-0.07}^{+0.11}$ & $-0.14_{-0.18}^{+0.11}$ & 1 \\ [2ex]
\hline
\hline
\captionsetup{singlelinecheck=false}
\end{tabular}
}
\end{center}
\vspace{-2mm}
\caption{ 
Data for the eight dSph and eight dIrr galaxies we study in this work. From left to right, the columns give: the name of the galaxy; type (dSph or dIrr); distance from the centre of the Milky Way; stellar mass; gas mass (for the dIrrs); stellar half light radius, $R_{1/2}$; exponential gas scale length (for the dIrrs); the pre-infall halo mass estimated from HI rotation curves (for the dIrrs) or abundance matching (for the dSphs; see \S\ref{sec:dmheating}); the number of kinematic member stars (for the dSphs); the star formation truncation time (defined in \S\ref{sec:intro}); and our estimates of $\rho_{\rm DM}(150\,{\rm pc})$ and $\gamma_{\rm DM}(150\,{\rm pc})$ with their 68\% confidence intervals (see \S\ref{sec:dmheating}). For the dSphs, the column giving $R_{1/2}$ quotes literature values compiled in the \citet{2012AJ....144....4M} review and, in square brackets, the value favoured for our sample of RGB stars by \GravSphere. This is in excellent agreement with the literature values for all dSphs except Sextans and UMi, where \GravSphere\ favours a smaller and larger $R_{1/2}$, respectively. Finally, the last column gives the data references for each galaxy, as follows: 1: \citet{2017MNRAS.467.2019R}; 2: \citet{2000ApJ...531..804D}; 3: \citet{2012AJ....144....4M}; 4: \citet{2001AJ....122.2524A}; 5: \citet{2002AJ....123.3199C}; 6: \citet{2012A&A...539A.103D}; 7: \citet{2009ApJ...703..692L}; 8: \citet{2002MNRAS.332...91D}; 9: \citet{2014A&A...572A..10D}; 10: \citet{2012A&A...544A..73D}. The references for the photometric and kinematic data for the dSphs are given in \S\ref{sec:data}.}
\label{tab:data}
\end{table*}

\section{Results}\label{sec:results}

\subsection{Example \GravSphere\ model fits and constraints on the velocity anisotropy profile}\label{sec:anisotropy}

\begin{figure*}
\begin{center}
\includegraphics[width=0.99\textwidth]{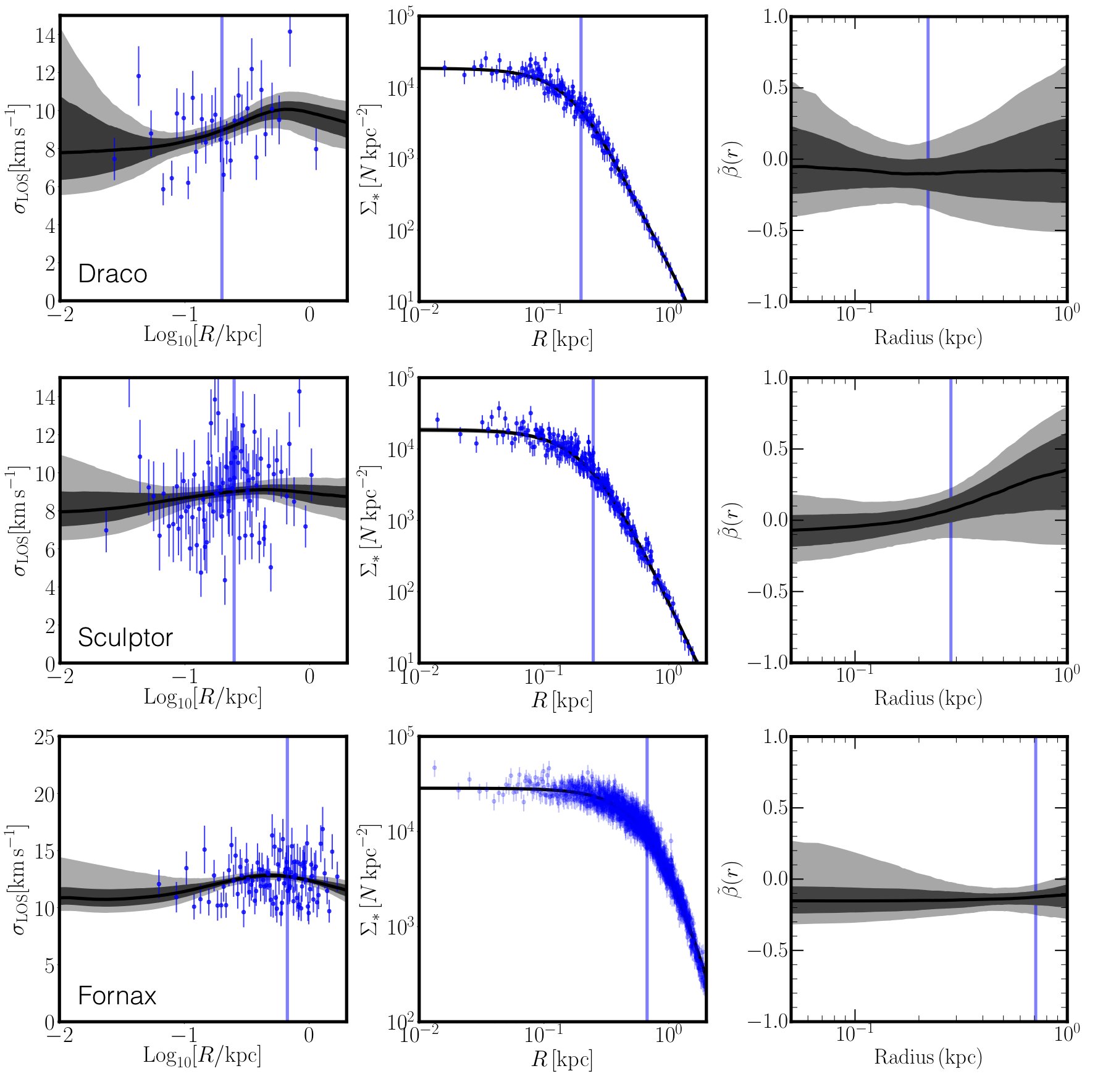}
\caption{Example \GravSphere\ model fits for Draco (top), Sculptor (middle) and Fornax (bottom). The panels show, from left to right, the projected velocity dispersion $\sigma_{\rm LOS}$, the tracer surface density profile, $\Sigma_*$ and the symmetrised velocity anisotropy profile, $\betastar$ (see equation \ref{eqn:betastar}). The data with errors are shown by the blue points, the contours mark the 68\% (dark grey) and 95\% (light grey) confidence intervals of our \GravSphere\ models, and the vertical blue lines mark the projected half light radius, $R_{1/2}$. From top to bottom, these three dSphs demonstrate the effect of increasing the number of member velocities, from 504 in Draco to 1,351 in Sculptor and 2,573 in Fornax. Notice how the constraints on $\betastar$ improve with improved spectroscopic sampling.}
\label{fig:dra_scl_for_fit}
\end{center}
\end{figure*}

Before addressing the primary goal of this work -- the DM density profiles -- in Figure \ref{fig:dra_scl_for_fit}, we show three example \GravSphere\ model fits for Draco (top), Sculptor (middle) and Fornax (bottom). (The other dSph fits are similar to these and so we omit them for brevity.) The panels show, from left to right, the projected velocity dispersion $\sigma_{\rm LOS}$, the tracer surface density profile, $\Sigma_*$ and the symmetrised velocity anisotropy profile, $\betastar$ (see equation \ref{eqn:betastar}). The data with errors are shown by the blue points, the contours mark the 68\% (dark grey) and 95\% (light grey) confidence intervals of our \GravSphere\ models, and the vertical blue lines mark the projected half light radius, $R_{1/2}$.

The three dSphs in Figure \ref{fig:dra_scl_for_fit} have an increasing number of member velocities, from 504 in Draco to 1,351 in Sculptor and 2,573 in Fornax. This demonstrates how the \GravSphere\ model fits improve with increasing sampling. Notice that in all cases, the \GravSphere\ models provide good fits to the binned data. Both VSPs (see \S\ref{sec:gravsphere}) are also well-fit for all three dwarfs, with no indication of bias due to triaxiality (see \citealt{2017MNRAS.471.4541R} for a discussion of this). The Draco model fits are discussed in detail in a separate companion paper where we use Draco -- that is the densest of our full dwarf sample -- to place constraints on SIDM models \citep{Read:2018pft}. 

For all of the dSphs that we study in this work, our \GravSphere\ models are consistent with being isotropic within their 95\% confidence intervals. The majority have strong constraints only near $R_{1/2}$ (c.f. the results for Draco in the top right panel of Figure \ref{fig:dra_scl_for_fit}). However, for Sculptor and Fornax, that have the largest number of member velocities, we are able to constrain $\betastar$ also at larger and smaller radii. For Sculptor, we weakly favour isotropic models near the centre that become radially anisotropic for $R > R_{1/2}$ (see Figure \ref{fig:dra_scl_for_fit}, middle row, right panel). For Fornax, we weakly favour some tangential anisotropy at all radii (see Figure \ref{fig:dra_scl_for_fit}, bottom row, right panel). Tangential anisotropy has been noted in some previous studies of Fornax \citep[e.g.][]{2013MNRAS.433.3173B,2018arXiv180707852K}. However, for our \GravSphere\ models, the evidence for this anisotropy is marginal.

\subsection{Dark matter density profiles}\label{sec:dmprofiles}

\begin{figure*}
\begin{center}
\includegraphics[width=0.99\textwidth]{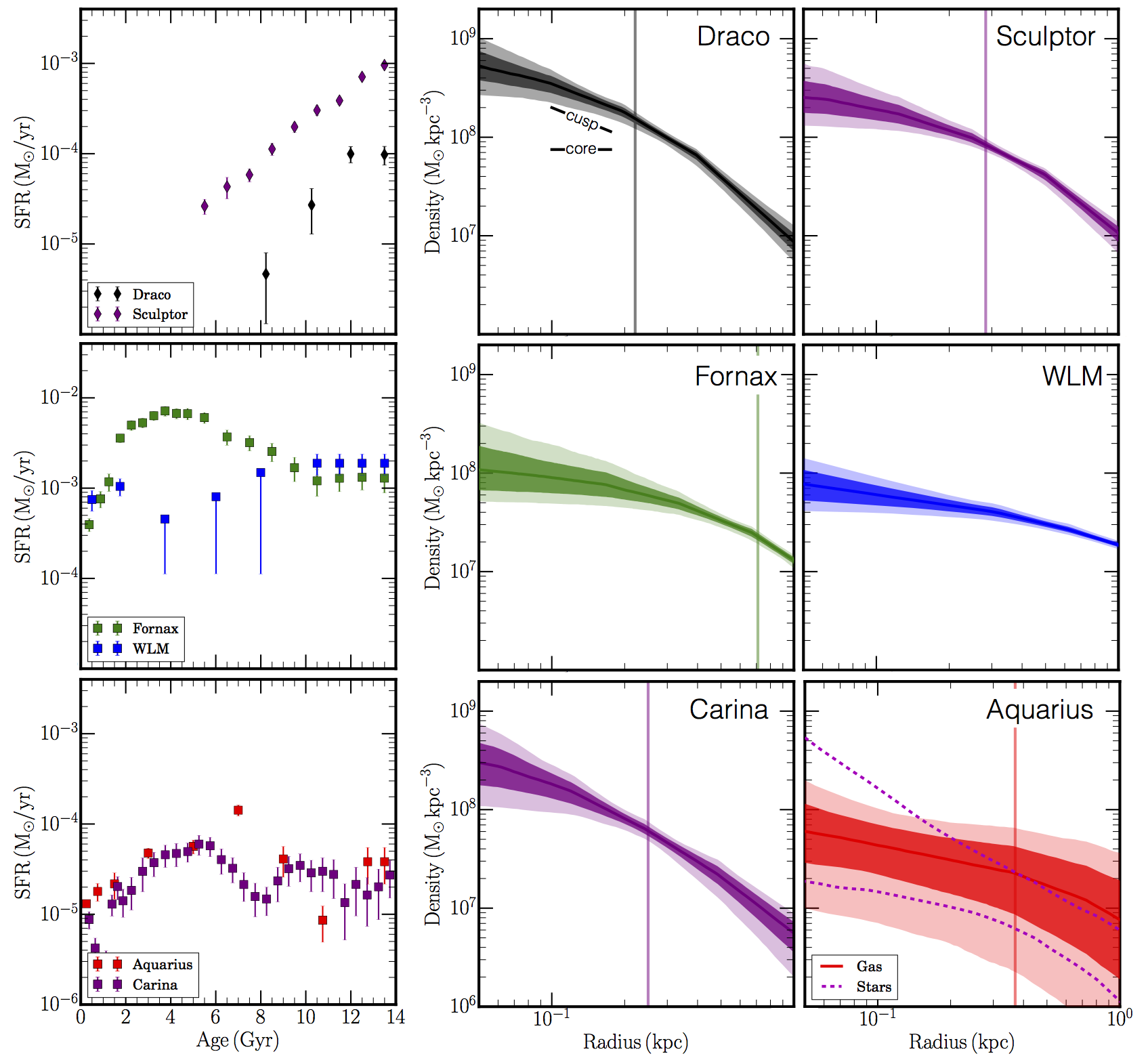}
\caption{The radial DM density profiles of dSphs with $>500$ member velocities, and two dIrrs (WLM and Aquarius) with a well-measured SFH (see \S\ref{sec:data}). The left panel shows the SFH, where today is on the left, while the beginning of the Universe is on the right of the plot. All plots are normalised such that the integral of the star formation rate over $t_{\rm univ} = 13.8$\,Gyrs matches the stellar masses reported in Table \ref{tab:data}. The middle and right panels show the radial DM density profiles. The light and dark contours mark the 95\% and 68\% confidence intervals of our models, respectively. The vertical grey lines mark the projected half light radius, $R_{1/2}$. For the dSphs, the DM density profile is derived from the stellar kinematics (\S\ref{sec:gravsphere}), while for the dIrrs it is derived from the HI gaseous rotation curve (\S\ref{sec:HIrot}). For Aquarius, there are also stellar radial velocities available for $\sim 25$ member stars. The purple dashed lines mark the 68\% confidence intervals of \GravSphere\ models applied to these data.}
\label{fig:killer_plot}
\end{center}
\end{figure*}

In Figure \ref{fig:killer_plot}, we show our results for the radial DM density profiles of dSphs with $>500$ member velocities, and two dIrrs -- WLM and Aquarius -- that have a well-measured SFH (see \S\ref{sec:data}). The left panel shows the SFH, where an age of zero corresponds to today, while the beginning of the Universe is on the right of the plot at $\sim 14$\,Gyrs. All plots are normalised such that the integral of the star formation rate over $t_{\rm univ} = 13.8$\,Gyrs matches the stellar masses reported in Table \ref{tab:data}. The middle and right panels show the radial DM density profiles. The light and dark contours mark the 95\% and 68\% confidence intervals of our models, respectively. The vertical grey lines mark the projected half light radius, $R_{1/2}$. For the dSphs, the DM density profile is derived from the stellar kinematics (\S\ref{sec:gravsphere}), while for the dIrrs it is derived from the HI gas rotation curve (\S\ref{sec:HIrot}). For Aquarius, there are also stellar radial velocities available for $\sim 25$ member stars\footnote{Note that stellar kinematic data are also available for WLM \citep{2012ApJ...750...33L}. However, there is evidence for rotation in these stars which cannot currently be included in the \GravSphere\ models. We will revisit joint constraints from combined stellar and gas kinematics in future work.} \citep{2014MNRAS.439.1015K}. The purple dashed lines mark the 68\% confidence intervals of \GravSphere\ models applied to these data. This demonstrates the consistency between our stellar kinematic and HI gas mass modelling, but -- as anticipated from tests on mock data in \citet{2017MNRAS.471.4541R} -- with just 25 member velocities, \GravSphere\ is not able to well-constrain the DM density inside $R < R_{1/2}$ for Aquarius.

\begin{figure*}
\begin{center}
\includegraphics[width=0.99\textwidth]{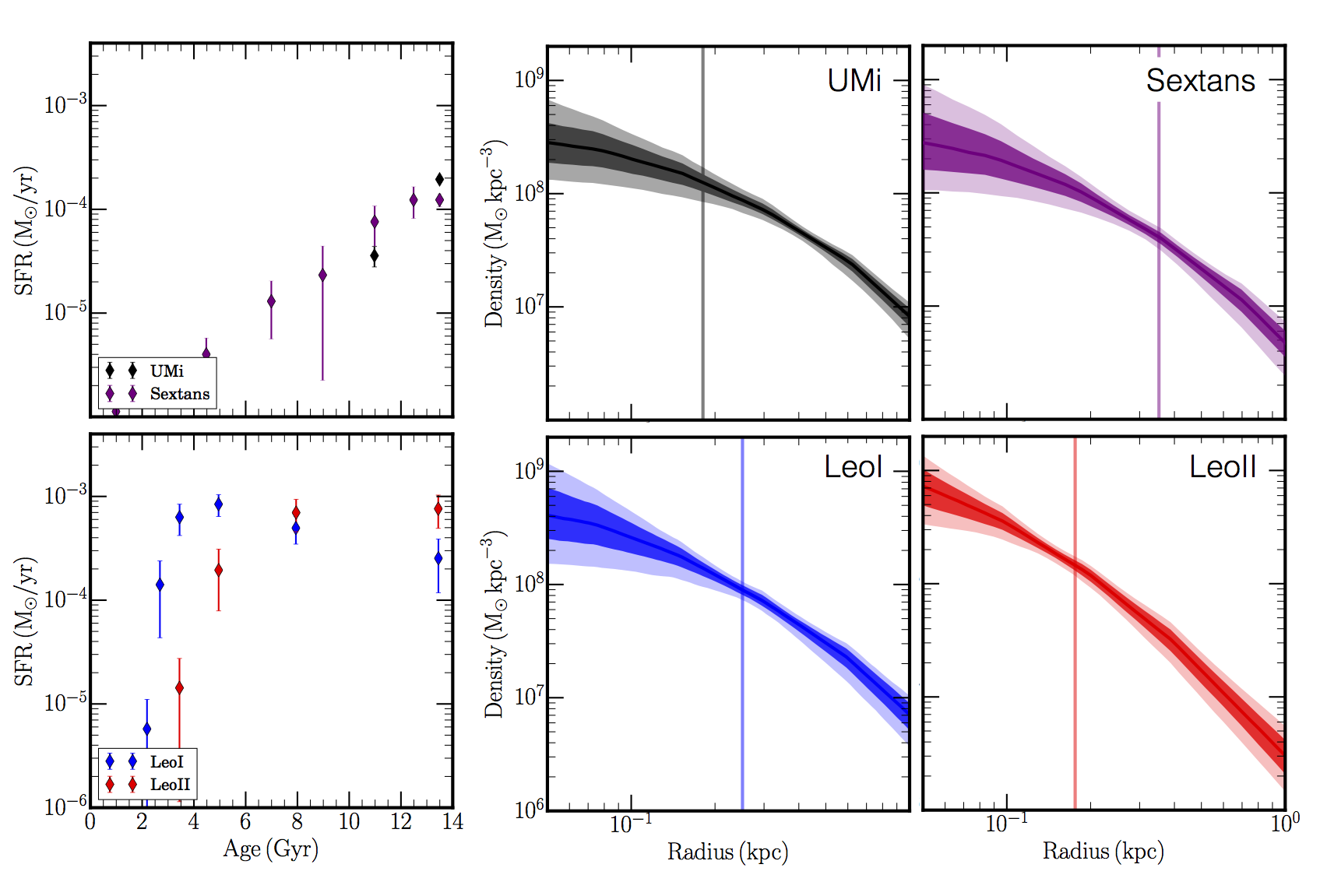}
\caption{As Figure \ref{fig:killer_plot}, but for dSphs with $<500$ member velocities.}
\label{fig:killer_plot_umisex}
\end{center}
\end{figure*}

Firstly, notice that the \GravSphere\ models for Draco favour a high central density inside $R < R_{1/2}$, consistent with a $\Lambda$CDM cusp. Below the contours of the \GravSphere\ models, we mark on two power law density profiles, $\rho \propto r^{-1}$ (cusp) and $\rho = {\rm const.}$ (core). (We discuss Draco, the densest dwarf of our full sample, in detail in a companion paper \citep{Read:2018pft}.) The \GravSphere\ models for Sculptor, that formed $\sim 8$ times more stars than Draco, favour a lower central density than Draco, consistent with both an inner core and a cusp within \GravSphere's 95\% confidence intervals. This trend of decreasing inner density with increasing star formation is seen also in Fornax. The \GravSphere\ models for Fornax -- that formed nearly 150 times more stars than Draco -- is less dense than both Draco and Sculptor, with $\rho_{\rm DM}(150\,{\rm pc})$ a factor of $\sim 3$ lower than for Draco. This shallow inner density profile for Fornax is remarkably similar to that for WLM (compare the middle and right panels in the middle row of Figure \ref{fig:killer_plot}). This is interesting since WLM and Fornax share similar SFHs (see Figure \ref{fig:killer_plot}, middle row, left panel) up until $\sim 2$\,Gyrs ago when Fornax's star formation quenched. Our \GravSphere\ models for Aquarius, despite having substantially larger uncertainties than WLM, also favour a low inner DM density within their 95\% confidence intervals. Finally, Carina is an interesting case. It has formed stars for nearly a full Hubble time, but despite its substantially more extended star formation, it formed only ${\sim}$30\% more stars than Draco. Our \GravSphere\ models for Carina weakly favour a dense `cuspy' profile, similar to that for Draco, but also permit a low density core within their 95\% confidence intervals (see Figure \ref{fig:killer_plot}, bottom row). We discuss Carina further in \S\ref{sec:discussion}.

In Figure \ref{fig:killer_plot_umisex}, we show the results for our sample of dSphs with $<500$ member velocities. For these galaxies, we expect the \GravSphere\ model constraints to be poorer and in general the confidence intervals of our models are broader for these dSphs. Nonetheless, we remain able to detect that Leo I and Leo II are substantially more dense than Fornax, while Sextans and UMi favour a density similar to Sculptor and Carina that lies in-between that of Draco and Fornax. Finally, in Appendix \ref{app:dIrrs} (Figure \ref{fig:killer_plot_dIrrs}), we show the results for the remainder of our sample of dIrrs. These all have similar DM density profiles that are consistent with constant density cores, as has been reported previously in the literature \citep[e.g.][]{2015AJ....149..180O,2017MNRAS.467.2019R}.

\subsection{A diversity of central dark matter densities}\label{sec:dmdiverse}

\begin{figure*}
\begin{center}
\includegraphics[width=0.975\textwidth]{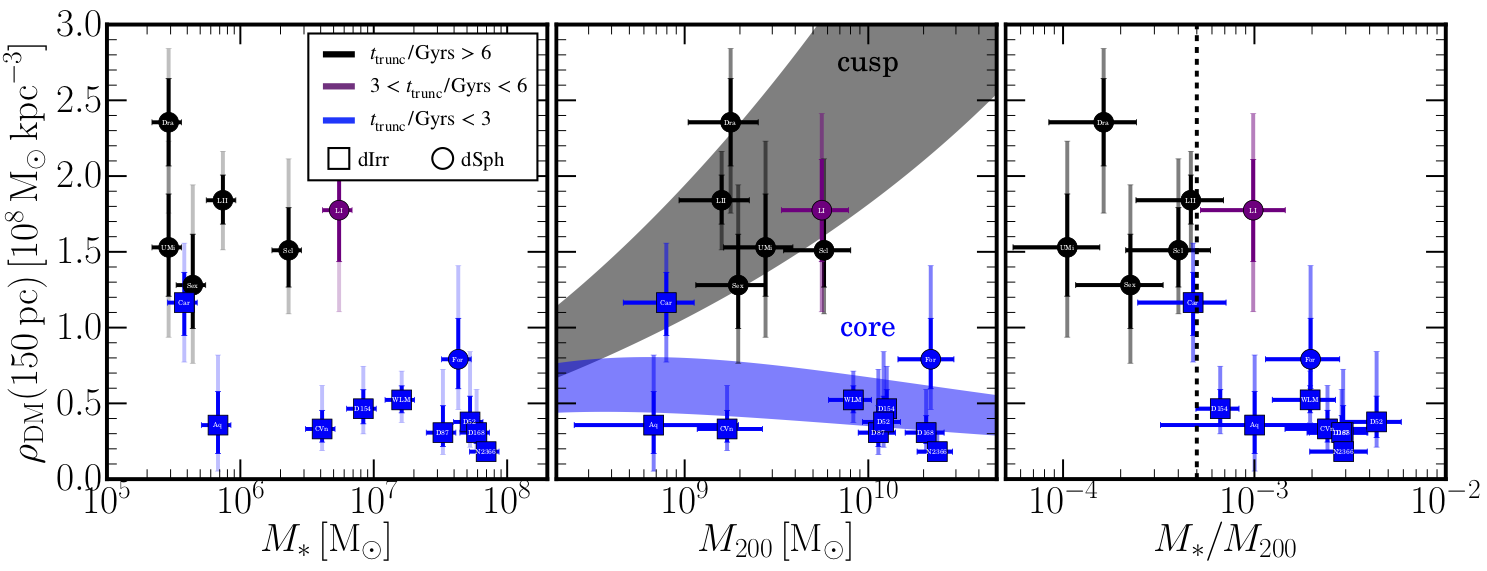}
\caption{{\bf Left:} The inner DM density of our sample of dwarfs, $\rho_{\rm DM}(150\,{\rm pc})$, as a function of their their stellar masses, $M_*$. The blue points mark those dwarfs that stopped forming stars $t_{\rm trunc} < 3$\,Gyrs ago; the black points those with $t_{\rm trunc} > 6$\,Gyrs; and the purple points those with $3 < t_{\rm trunc}/{\rm Gyrs} < 6$ (see Table \ref{tab:data}). The square symbols denote dIrr galaxies, whose central densities were determined from their HI rotation curves (\S\ref{sec:HIrot}); the circle symbols denote dSph galaxies, whose central densities were determined from their stellar kinematics (\S\ref{sec:gravsphere}). Notice that dwarfs with extended star formation (blue) have $\rho_{\rm DM}(150\,{\rm pc}) < 10^8$\,M$_\odot$\,kpc$^{-3}$, while those with only old stars (black) have $\rho_{\rm DM}(150\,{\rm pc}) > 10^8$\,M$_\odot$\,kpc$^{-3}$. Notice also the `dwarf twins' -- UMi, Draco, Carina, Sextans, Leo II and Aquarius -- that have similar $M_*$ but very different $\rho_{\rm DM}(150\,{\rm pc})$. {\bf Middle:} $\rho_{\rm DM}(150\,{\rm pc})$ as a function of pre-infall halo mass, $M_{200}$, as extrapolated from HI rotation curves (for the dIrrs) and abundance-matching (for the dSphs; see \S\ref{sec:data}). The grey band marks the inner DM density of $\Lambda$CDM halos assuming no cusp-core transformations take place, where the width of the band corresponds to the $1\sigma$ scatter in DM halo concentrations (equation \ref{eqn:M200c200}). The blue band marks the same, but for the \coreNFW\ profile from \citet{2016MNRAS.459.2573R}, assuming maximal core formation. Thus, these two bands bracket the extremum cases of no cusp-core transformation and complete cusp-core transformation in $\Lambda$CDM. Notice that dwarfs with extended star formation (blue) lie along the blue track, consistent with having DM cores, while those whose star formation shut down long ago (black) lie along the grey track, consistent with having DM cusps. {\bf Right:} $\rho_{\rm DM}(150\,{\rm pc})$ as a function of the stellar mass to halo mass ratio, $M_*/M_{200}$. Notice that dwarfs that have formed more stars as a fraction of their pre-infall halo mass have a lower central dark matter density. This is consistent with models in which DM is `heated up' by repeated gas inflows and outflows driven by stellar feedback \citep[e.g.][]{2005MNRAS.356..107R,2012MNRAS.421.3464P,2014MNRAS.437..415D,2015MNRAS.454.2981C,2016MNRAS.456.3542T}. The vertical dashed line marks the approximate $M_*/M_{200}$ ratio below which recent models predicted that DM cusp-core transformations should become inefficient \citep{2012ApJ...759L..42P,2014MNRAS.437..415D,2015MNRAS.454.2981C,2016MNRAS.456.3542T}.}
\label{fig:dmheating}
\end{center}
\end{figure*}

In Figures \ref{fig:killer_plot} and \ref{fig:killer_plot_umisex}, we saw that our sample of dwarfs have a wide range of dark matter density profiles. In particular, their central densities appeared to decrease with increasing star formation. In this section, we study this diversity quantitatively. In Figure \ref{fig:dmheating}, left panel, we plot $\rho_{\rm DM}(150\,{\rm pc})$ for our full sample of dwarfs (see \S\ref{sec:cuspcoretheory}) as a function of their stellar masses, $M_*$. The data points are coloured by their star formation truncation times, $t_{\rm trunc}$, as marked in the legend (see \S\ref{sec:data} and Table \ref{tab:data}). Notice that the dwarfs fall into two broad classes. Those with only old stars ($t_{\rm trunc} > 6$\,Gyrs; black) have $\rho_{\rm DM}(150\,{\rm pc}) > 10^8$\,M$_\odot$\,kpc$^{-3}$, while those with extended star formation ($t_{\rm trunc} < 3$\,Gyrs; blue) have $\rho_{\rm DM}(150\,{\rm pc}) < 10^8$\,M$_\odot$\,kpc$^{-3}$. Note, however, that Carina, UMi and Sextans are possible exceptions to this. They could lie on either side of this boundary within their 95\% confidence intervals. This could imply a continuum of central dark matter densities rather than a dichotomy. However, the uncertainties on $\rho_{\rm DM}(150\,{\rm pc})$ are currently too large to determine whether or not this is the case. We discuss this further in \S\ref{sec:discussion}.

Finally, notice that there are several dwarfs -- UMi, Draco, Carina, Sextans, Leo II and Aquarius -- with similar baryonic mass but very different $\rho_{\rm DM}(150\,{\rm pc})$. In Appendix \ref{app:dwarftwins}, we show that this is challenging to understand in `alternative gravity' theories for DM.

\subsection{Evidence for dark matter heating in dwarf galaxies}\label{sec:dmheating}

From Figure \ref{fig:dmheating}, left panel, we see a significant scatter in the central DM densities of nearby dwarf galaxies at a similar stellar mass. In this section, we consider three physical effects that could induce this scatter in $\Lambda$CDM. Firstly, ram pressure from the Milky Way's hot corona will cause star formation in the dwarfs to rapidly shut down on infall \citep[e.g.][]{2013MNRAS.433.2749G}. This will induce scatter in $M_*$ at a fixed pre-infall halo mass, $M_{200}$, leading to a range of $M_*$ at a given $\rho_{\rm DM}(150\,{\rm pc})$ \citep[e.g.][]{2017MNRAS.467.2019R}. Secondly, tidal shocking and stripping can lower the central DM density of the dwarfs, inducing scatter in $\rho_{\rm DM}(150\,{\rm pc})$ at a fixed $M_*$ \citep[e.g.][]{2003ApJ...584..541H,2004ApJ...608..663K,2006MNRAS.366..429R}. And thirdly, `DM heating' will push dark matter out of the centres of the dwarfs. At fixed $M_{200}$, this leads to a lower $\rho_{\rm DM}(150\,{\rm pc})$ for a larger $M_*$ (see \S\ref{sec:intro}).

Firstly, note that while tidal stripping is likely to affect the outer dark matter profiles of the dSphs, for the orbits that the classical dwarfs are known to move on, the effect of tidal stripping and shocking on the profile inside $R_{1/2}$ is expected to be small \citep[e.g.][]{2003ApJ...584..541H,2004ApJ...608..663K,2006MNRAS.366..429R,2006MNRAS.367..387R,2008ApJ...673..226P,2010MNRAS.406.2312L,Read:2018pft,2018arXiv180409381G}. Furthermore, tides cannot affect the isolated dIrrs, yet these have a {\it lower} $\rho_{\rm DM}(150\,{\rm pc})$ than most of the dSphs (Figure \ref{fig:dmheating}, left panel). Of the mechanisms we consider here, this leaves ram pressure stripping and DM heating as the main sources of scatter.

Ram pressure-induced scatter in $\rho_{\rm DM}(150\,{\rm pc})$ at a fixed $M_*$ is caused, ultimately, by the dwarfs inhabiting halos with very different pre-infall masses, $M_{200}$. Thus, if we can obtain an independent estimate of $M_{200}$ for our dwarf sample, then we can remove this source of scatter. As discussed in \S\ref{sec:data}, for the isolated dIrrs we obtain an extrapolated $M_{200}$ directly from their HI rotation curves \citep{2017MNRAS.467.2019R}, while for the dSphs, we obtain $M_{200}$ by abundance matching with their mean star formation rates \citep{2018arXiv180707093R}. We report these $M_{200}$ for our full sample, with uncertainties, in Table \ref{tab:data}.

In Figure \ref{fig:dmheating}, middle panel, we plot $\rho_{\rm DM}(150\,{\rm pc})$ as a function of $M_{200}$ for our full dwarf sample. The grey band marks the expected range of inner DM densities of $\Lambda$CDM halos assuming no cusp-core transformations take place (i.e. assuming NFW profiles), where the width of the band accounts for the 1$\sigma$ scatter in the $M_{200}-c_{200}$ relation (see equations \ref{eqn:rhoNFW}, \ref{eqn:M200c200} and Figure \ref{fig:NFW_similar}). The blue band marks the same, but for the \coreNFW\ profile, assuming maximal core formation (equation \ref{eqn:coreNFW}). Thus, the grey and blue bands bracket the extremum cases of no cusp-core transformation and complete cusp-core transformation in $\Lambda$CDM. 

From Figure \ref{fig:dmheating}, middle panel, we can see that the dwarfs with extended star formation (blue) have low central DM densities and lie along the blue track, consistent with DM cores, while those whose star formation shut down long ago (black) lie along the grey track, consistent with DM cusps. The uncertainties on $\rho_{\rm DM}(150\,{\rm pc})$ and $M_{200}$ are currently too large to be able to definitively place any of the dwarfs in the `transition' region between being fully cusped (grey) and fully cored (blue). We discuss this further in \S\ref{sec:discussion}.

To further illustrate the above result, in Figure \ref{fig:dmheating}, right panel, we plot $\rho_{\rm DM}(150\,{\rm pc})$ for our sample of dwarfs as a function of the ratio of their stellar mass, $M_*$ to their pre-infall halo mass, $M_{200}$. Now the anti-correlation between star formation and the central DM density is explicit: dwarfs with higher $M_*/M_{200}$ have lower $\rho_{\rm DM}(150\,{\rm pc})$. This is in excellent agreement with models in which DM is heated up by bursty star formation. Several works in the literature, using different numerical techniques and different `sub-grid' star formation recipes, predicted that DM cusp-core transformations should become inefficient\footnotemark\ for $M_*/M_{200} \simlt 5 \times 10^{-4}$ (\citealt{2012ApJ...759L..42P,2014MNRAS.437..415D,2015MNRAS.454.2981C,2016MNRAS.456.3542T}, and for a review see \citealt{2017ARA&A..55..343B}). This is marked by the vertical dashed line on Figure \ref{fig:dmheating}, right panel. Notice, further, that this line delineates dwarfs that have extended star formation (blue) from those with only old-age stars (black).

\footnotetext{Note that \citet{2014MNRAS.437..415D} actually set this boundary to be $M_*/M_{200} = 10^{-4}$. However, from their Figure 3, this corresponds to $\gamma_{\rm DM}(0.01 < r/R_{200}<0.02) \sim -1$. At $M_*/M_{200} = 5 \times 10^{-4}$, most of their simulations still have $\gamma_{\rm DM}(0.01 < r/R_{200}<0.02) \sim -0.85$, corresponding to very little cusp-core transformation.}

Finally, recall that for $\sim 500$ radial velocities, \GravSphere's inference of the inner logarithmic slope of the DM density profile -- $\gamma_{\rm DM}(150\,{\rm pc}) \equiv d\ln\rho_{\rm DM}/d\ln r(150\,{\rm pc})$ -- depends on our choice of priors on $\gamma_{\rm DM}$ \citep{Read:2018pft}. For this reason, we have focussed in this paper only on the amplitude of the inner DM density, $\rho_{\rm DM}(150\,{\rm pc})$ (see \S\ref{sec:cuspcoretheory}). Nonetheless, for completeness we show our results for $\gamma_{\rm DM}(150\,{\rm pc})$ in Appendix \ref{app:gammaDM}. There, we confirm that $\gamma_{\rm DM}(150\,{\rm pc})$ is sensitive to our priors on $\gamma_{\rm DM}$. However, independently of our priors on $\gamma_{\rm DM}$, we find that dwarfs with truncated star formation have {\it steeper} central density profiles than those with extended star formation, consistent with our results for $\rho_{\rm DM}(150\,{\rm pc})$, above.

We have shown that the scatter in $\rho_{\rm DM}(150\,{\rm pc})$ at fixed $M_*$ (Figure \ref{fig:dmheating}, left panel) cannot owe to tidal stripping and shocking. Tidal effects are certainly important for some of the Milky Way dwarfs (for example the visibly disrupting Sagittarius dSph; \citealt{1995MNRAS.277..781I}). However, the sample of dSphs that we have considered in this paper are moving on relatively benign orbits around the Milky Way. Their orbits are not sufficiently radial to affect the DM density at 150\,pc \citep[e.g.][]{2010MNRAS.406.2312L,2018arXiv180409381G}. We have shown further that the scatter cannot owe to the dwarfs inhabiting different pre-infall mass halos. The dwarfs certainly do inhabit a range of different pre-infall halo masses (Figure \ref{fig:dmheating}, middle panel). However, this is not sufficient to explain the scatter we find in $\rho_{\rm DM}(150\,{\rm pc})$. In particular, we see no correlation between $\rho_{\rm DM}(150\,{\rm pc})$ and $M_{200}$ (Figure \ref{fig:dmheating}, middle panel). By contrast, we see a clear anti-correlation between $\rho_{\rm DM}(150\,{\rm pc})$ and the ratio $M_*/M_{200}$ (Figure \ref{fig:dmheating}, right panel). This anti-correlation was predicted by models in which DM is slowly `heated up' at the centres of dwarf galaxies by bursty star formation \citep{2012ApJ...759L..42P,2014MNRAS.437..415D,2015MNRAS.454.2981C,2016MNRAS.459.2573R,2016MNRAS.456.3542T}. In \S\ref{sec:discussion}, we discuss which combination of measurements would need to be wrong in order for this agreement between data and models to be spurious.

\section{Discussion}\label{sec:discussion}

%MOREMORE CHECK THIS SECTION FOR CLARITY WRT SFH => HEATING 

\subsection{Comparison with previous work in the literature}\label{sec:compare}

\subsubsection{The dwarf irregulars}

Our sample of dIrrs is drawn from the Little THINGS survey \citep{2015AJ....149..180O,2017MNRAS.466.4159I}. \citet{2015AJ....149..180O} presented mass models for all of the dIrrs we discuss here, using an entirely independent derivation of their rotation curves from the raw HI datacubes. \citet{2015AJ....149..180O} also favour DM cores for these dIrrs, finding an inner logarithmic slope, averaged over their full sample, of $\gamma_{\rm DM} = -0.32 \pm 0.24$. This is in excellent agreement with our findings here (see Table \ref{tab:data}). The only dIrrs for which \citet{2015AJ....149..180O} favour DM cusps are DDO 101 and DDO 210 (Aquarius). DDO 101 was discussed extensively in \citet{2016MNRAS.462.3628R}. There, it was shown that DDO 101's steeply rising rotation curve could owe to an incorrect distance estimate for this dwarf. Indeed, DDO 101 did not make our final selection precisely because of its highly uncertain distance. For DDO 210, we find, similarly to \citet{2015AJ....149..180O}, that the uncertainties on the inner DM density and logarithmic slope are simply very large (see Table \ref{tab:data}). In terms of the inner logarithmic slope of its DM density profile, Aquarius could indeed be cusped or cored within its 95\% confidence intervals (see Figure \ref{fig:dmheating}). However, the {\it amplitude} of Aquarius' inner DM density, $\rho_{\rm DM}(150\,{\rm pc})$, is consistent with it being cored (see Figure \ref{fig:dmheating}, right panel).

\subsubsection{The dwarf spheroidals}  

Among the dSphs, by far the most well-studied are Fornax and Sculptor, which are relatively luminous and have the largest available stellar-kinematic samples (for reviews, see \citealt{2013NewAR..57...52B} and \citealt{2013pss5.book.1039W}). While there is a general consensus that Fornax has a DM core \citep{GoerdtEtAl2006,2011ApJ...742...20W,2011MNRAS.tmp.1606A,2012MNRAS.426..601C,2018arXiv180202606P,2018arXiv180707852K}, Sculptor has proven more contentious.  For example, modeling split populations using the Jeans equations and/or the Virial theorem, \citet{2008ApJ...681L..13B}, \citet{2012ApJ...754L..39A}, \citet{2011ApJ...742...20W} and \citet{2012MNRAS.419..184A} all favour a central DM core; using VSPs similar to our analysis here, \citet{2014MNRAS.441.1584R} favour a cusp; using a Schwarzschild method, split-populations with axisymmetric Jeans models and a phase-space distribution function method, respectively, \citet{2013MNRAS.433.3173B}, \citet{2016MNRAS.463.1117Z} and \citet{2017ApJ...838..123S} all conclude that they cannot distinguish cusps from cores with the currently-available data. Finally, \citet{2017arXiv171108945M} have recently used the first internal proper motion data for Sculptor to argue that it favours a cusp. However, \citet{2018arXiv180107343S} argue that those same proper motion data are consistent with both cusps and cores.

\begin{figure*}
\begin{center}
\includegraphics[width=0.99\textwidth]{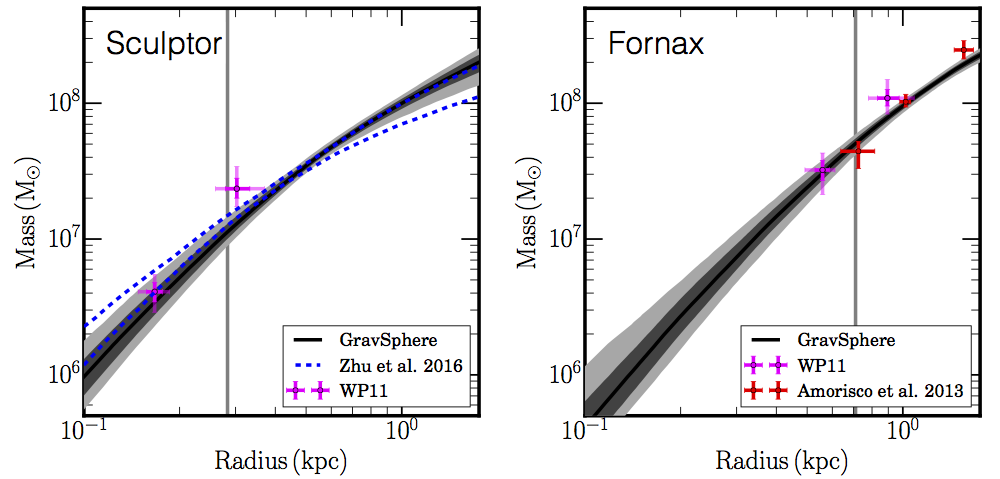}
\caption{The cumulative DM mass profile of our \GravSphere\ models for Sculptor (left) and Fornax (right) as compared to other determinations in the literature (see legend). The grey contours show the 68\% and 95\% confidence intervals of our \GravSphere\ models.}
\label{fig:compare}
\end{center}
\end{figure*}

Figure \ref{fig:compare} compares our new results for the cumulative DM mass profiles of Sculptor (left) and Fornax (right) to those from previous studies for which such a comparison is straightforward\footnote{Previous studies that we have not included in this plot evaluate perfectly cored and/or NFW-cusped halo models separately. This makes it challenging to compare with our \GravSphere\ models that provide a posterior probability distribution function that includes the space in between these two extremes.}. The grey contours show the 68\% (dark) and 95\% (light) confidence intervals of our \GravSphere\ models. The magenta and red data points show the results from \citet{2011ApJ...742...20W} and \citet{2013MNRAS.429L..89A}, respectively, who both use split population methods with dynamical mass estimators to obtain measurements of the enclosed masses at different scale radii. (The light/dark error bars mark the 95\% and 68\% confidence intervals of these models, respectively.) The dashed blue curves indicate the posterior PDF that \citet{2016MNRAS.463.1117Z} obtain for a generalized DM halo model, using split populations with an axisymmetric Jeans method that includes rotation.  All of these methods break the $\rho-\beta$ degeneracy (see \S\ref{sec:intro}) in different ways, while each study uses their own data selection and their own approach to determining the membership probability.

Most of the mass models for Sculptor and Fornax shown in Figure \ref{fig:compare} agree within their 68\% confidence intervals. This is remarkable given the different methodologies used to derive these mass profiles. However, a notable outlier is the Sculptor result of \citet[][WP11 hereafter]{2011ApJ...742...20W}, who report an enclosed mass at $r\sim 300$ pc that is a factor of $\sim 2$ larger than that obtained in the other studies (including the present one). It is this large mass -- or more precisely, the relatively steep slope required to reach this mass from WP11's more-agreeable estimate at smaller radius -- that leads WP11 to conclude that Sculptor's mass profile is incompatible with an NFW cusp.  WP11's methodology has been tested extensively using mock data sets drawn from equilibrium dynamical models as well as cosmological and hydrodynamical N-body simulations, generally supporting WP11's argument that it is the mass at \textit{smaller} radius that is more prone to overestimation \citep[e.g.][]{2013MNRAS.433L..54L,2017arXiv170706303G}. However, the outer mass can be overestimated in the case of ongoing tidal heating (see the discussion by WP11) and/or departures from spherical symmetry that can conspire with unfortunate viewing angles to bias WP11's mass estimator. Even so, \citet{2017arXiv170706303G} find that in just $\sim 3\%$ of their cosmologically-simulated realisations of Sculptor analogs with cuspy DM halos, the latter effect would induce sufficient systematic error to account for WP11's result. 

At present, we lack a satisfactory explanation for the apparent $\sim 2\sigma$ systematic discrepancy, above. However, the key result in this paper -- that we find an anti-correlation between $\rho_{\rm DM}(150\,{\rm pc})$ and $M_*/M_{200}$ -- is based on the inference of $\rho_{\rm DM}$ at 150\,pc where all of the above studies agree. Furthermore, the trend exhibited across the population of dwarf galaxies in our sample should be insensitive to even large systematic errors in the mass profiles inferred for individual systems, provided that the systematic errors do not correlate with the star formation history.

\subsubsection{Dark matter heating}

From the above comparisons, it is clear that the results in this paper do not owe to any special feature of our \GravSphere\ modelling. Rather, what is new here is: (i) the comparison of the DM distribution in isolated gas rich dwarfs with our sample of nearby gas poor dwarf spheroidals; and (ii) the comparison of the inner DM density of these dwarfs with their SFHs. With a large sample of such dwarfs with excellent quality data, we are able to demonstrate that Fornax, with its extended star formation history, has a shallow DM density profile similar to that of WLM and the other isolated dIrrs, while nearby dSphs that have only old-age stars are substantially denser, consistent with steeper, more cuspy, DM density profiles. These results are in good agreement with recent predictions by \citet{2018arXiv180607679B} who used energetic arguments to show that UMi and Draco are the dSphs most likely to have a pristine DM cusps, while Fornax and Sculptor are most likely to have large DM cores. Similarly, \citet{2015MNRAS.450.3920B} used their DM heating models, combined with abundance matching, to predict DM cores in WLM and Fornax, cusps in Draco, Leo I, Leo II and UMi, and something in-between for Sculptor and Aquarius. This is also in excellent agreement with our findings here.

Finally, the diversity of central DM densities that we find here is in good agreement with the recent study of \citet{2017arXiv171103502V}. They fit a self interacting DM (SIDM) model to the classical dSphs, finding a wide range of interaction cross sections, corresponding to a wide range of central DM densities. Similarly to our results here, they favour a low central density (high SIDM cross section) for Fornax and a high central density (low SIDM cross section) for Draco\footnote{For the remaining dwarfs, our study and that of \citet{2017arXiv171103502V} are broadly in good agreement, though they claim tighter constraints on the central density for UMi and Sextans than our \GravSphere\ models are able to achieve.}. However, without the dIrrs to compare with, they describe Fornax (and Sextans) as `outliers'. We favour a different interpretation. Given the good agreement between the inferred DM density profile of Fornax and that of our dIrr sample, we argue that Fornax is not an outlier, but rather a key piece of evidence for DM heating at the centres of dwarf galaxies.

\subsection{Model limitations and caveats}\label{sec:caveats}

\subsubsection{Mass modelling with stellar kinematics}

In recent years, there have been a number of studies critiquing the robustness of stellar kinematic mass modelling. The primary concerns are the effects of unmodelled triaxiality and the effect of unbound tidally stripped stars. Four recent studies have looked at the effects of triaxiality on mass modelling methods that assume spherical symmetry. \citet{2017MNRAS.471.4541R} test the \GravSphere\ method that we use here; \citet{2013MNRAS.433L..54L} and \citet{2017arXiv170706303G} test the \citet{2011ApJ...742...20W} split-population method; and \citet{2017arXiv170809425K} test a Schwarzschild method. All four find that triaxiality induces a small bias on the recovery that is rarely larger than the 95\% confidence intervals of the models. \citet{2013MNRAS.431.2796K} test the \citet{2010MNRAS.406.1220W} Jeans mass estimators on tidally stripped mock data, finding that they can become significantly biased. This contrasts with our recent work in \citet{Read:2018pft} where we show that \GravSphere\ is able to successfully recover the radial density profile of a tidally stripped mock dwarf set up to mimic Draco, within its 95\% confidence intervals. A full exploration of this difference is beyond the scope of this present work, but may owe to \citet{2013MNRAS.431.2796K} using Jeans mass estimators that are more prone to bias than fully self-consistent dynamical models \citep[e.g.][]{2017MNRAS.469.2335C}, or to their mocks being further from dynamical equilibrium than those considered in \citet{Read:2018pft}.

\subsubsection{Mass modelling with HI rotation curves}
The list of potential pitfalls for modelling gaseous rotation curves is rather longer than for stellar kinematic mass modelling. Several studies have worried about the effects of beam smearing \citep[e.g.][]{2002ApJ...575..801M}, non-circular motions due to a central bar \citep[e.g.][]{2004ApJ...617.1059R,2007ApJ...657..773V}, unmodelled turbulent or vertical pressure support in the disc \citep[e.g.][]{2007ApJ...657..773V,2017MNRAS.466...63P}, inclination error \citep[e.g.][]{2004ApJ...617.1059R,2016MNRAS.462.3628R}, umodelled halo triaxaility \citep[e.g.][]{2006MNRAS.373.1117H,2011MNRAS.414.3617K,2017arXiv170607478O} and the effect of large HI holes driven by supernovae explosions \citep{2016MNRAS.462.3628R}. In \citet{2016MNRAS.462.3628R}, we explicitly tested the methodology we use here on high resolution mock rotation curve data that include most of the above potential problems. We found that for fitted inclinations $i > 40^\circ$ (which is the case for all of the galaxies we consider in this paper), we were able to successfully recover the underlying rotation curve and obtain the correct mass distribution. The only issue that we did not explore in \citet{2016MNRAS.462.3628R} is the effect of non-circular motions due to halo triaxiality or a stellar bar. None of the galaxies in the sample we use here has a prominent stellar or gaseous bar, but they could inhabit triaxial DM halos. \citet{2017arXiv170607478O} have recently argued that this could be a significant source of bias in rotation curve modelling that typically assumes, as we have done here, a spherical DM halo. They demonstrate, using mock data from the APOSTLE simulations, that non-circular motions due to halo triaxiality can cause DM cusps to masquerade as cores. However, the mock dwarf galaxies used in \citet{2017arXiv170607478O} appear to have significantly larger non-circular motions (as determined from the residuals of their tilted ring model fits) than real galaxies in the Little THINGS survey \citep{2015AJ....149..180O,2017MNRAS.466.4159I}. Furthermore, triaxiality should induce a range of apparent inner DM logarithmic cusp slopes, with some galaxies appearing cored and others appearing cusped. This is not what we find for our sample of dIrrs that all favour a constant density DM core (see Figures \ref{fig:dmheating} and \ref{fig:killer_plot_dIrrs}). Nonetheless, this is an issue that warrants more attention in future work.

\subsubsection{Systematic bias between stellar kinematic and HI rotation curve modelling}
Almost all of our high density dwarfs are gas-free dwarf spheroidals, while our low density dwarfs are all gas rich dwarf irregulars. This general trend is expected if DM is heated up by bursty stellar feedback \citep[e.g.][]{2014MNRAS.437..415D,2016MNRAS.459.2573R}. However, the dwarf spheroidals are modelled using stellar kinematics, while the dwarf irregulars are modelled using gaseous rotation curves. Could this modelling difference be the true cause of the density-dichotomy that we see here? To answer this question, it is instructive to consider two scenarios in which the results in Figure \ref{fig:dmheating} are spurious and owe to some problem with our mass modelling. In scenario A, let us suppose that all dwarfs are actually cusped, with a central density $\rho_{\rm DM}(150\,{\rm pc}) > 10^8$\,M$_\odot$\,kpc$^{-3}$. In this case, the following would have to be true: (i) all stellar kinematic studies to date have mis-measured Fornax's DM density profile (c.f. \S\ref{sec:compare}); (ii) Fornax's globular clusters have found some way, as yet unknown, to survive orbiting in a dense cusped DM halo \citep{GoerdtEtAl2006,2012MNRAS.426..601C}; (iii) the agreement between the inner DM density profile of Fornax derived using \GravSphere\ and the dwarf irregulars is an unfortunate coincidence (Figures \ref{fig:killer_plot} and \ref{fig:killer_plot_dIrrs}); and (iv) all of the dwarf irregular density profiles presented in this paper are wrong. In scenario B, let us suppose that all dwarfs have large cores of size $\simgt R_{1/2}$, with central densities $\rho_{\rm DM}(150\,{\rm pc}) < 10^8$\,M$_\odot$\,kpc$^{-3}$. In this case: (i) the \GravSphere\ density profiles for Draco, Sculptor, Leo I and Leo II are wrong; (ii) the remaining dSphs must lie at the 95\% lower bound of their \GravSphere\ model density profiles (Figure \ref{fig:dmheating}, middle panel); and (iii) \GravSphere\ works on mock data but fails on the real data for most dSphs. Both scenarios seem unlikely. While the results for any individual dwarf may change, it seems hard to escape the conclusion that some dwarfs have a high central DM density, while others have low central DM density. 

\subsubsection{Systematic uncertainties in the pre-infall halo masses}
The results in Figures \ref{fig:dmheating}, middle and right panels, rely on estimates for the pre-infall halo masses, $M_{200}$, of our dwarf sample. For the gas rich dIrrs, we took these from the HI rotation curve estimates in \citet{2017MNRAS.467.2019R}; for the dSphs, we used the abundance matching method from \citet{2018arXiv180707093R}. While both of these estimates could suffer from sizeable systematic uncertainties, such errors cannot explain the diversity of central dark matter densities that we find here. If we wanted all of the dwarfs to lie along the grey track in Figure \ref{fig:dmheating}, middle panel, we would have to have Fornax and all of the dIrrs inhabit halos with masses $M_{200} < 5 \times 10^8$\,M$_\odot$, inconsistent with the peak rotation curve measurements for our dIrr sample. Furthermore, we would not be able to explain how such low mass galaxies managed to form so many stars. For these reasons, we are confident that our results are not contingent on our pre-infall halo mass estimates.

\subsection{A dichotomy or a continuum of cusps and cores?}

At present, our results in Figures \ref{fig:dmheating} are consistent with some dwarfs being cusped (those with only old-age stars; black), and some dwarfs being cored (those with younger stars; blue). However, as the constraints on $\rho_{\rm DM}(150\,{\rm pc})$ improve, we may find galaxies in transition between being fully cusped or fully cored. Leo I, with a star formation truncation time of $t_{\rm trunc} = 3.1$\,Gyrs, is a good candidate for such a dwarf, frozen in transition. Furthermore, we may find that the correspondence between being cusped or cored and $t_{\rm trunc}$ is not exact (see \S\ref{sec:intro}). There could be significant stochasticity in the formation of DM cores, driven by differing merger histories \citep[e.g.][]{2015MNRAS.449L..90L} and/or the spin and concentration parameters of the dwarfs' dark matter halos \citep[e.g.][]{2016MNRAS.459.2573R}. Carina is particularly interesting in this regard as it has an extended star formation history, yet weakly favours a dark matter cusp (Figure \ref{fig:dmheating}). Similarly, the `ultra-faint' dwarfs Eridanus II and Andromeda XXV may be further examples of stochasticity, since both appear to have old-age stars and central DM cores \citep{2017ApJ...844...64A,2017arXiv170501820C}. (Note, however, that the cores claimed in these ultra-faint dwarfs are much smaller than the $\sim 150$\,pc scale that we are able to probe here. As such, an alternative explanation could be that all dSphs -- both classical and ultra-faint -- have a small $\simlt 100$\,pc-size inner core that forms at high redshift, and that we are not yet able to detect yet. See \citet{Read:2018pft} for some further discussion on this point.) We will address these questions in more detail in future work.

\subsection{The Too Big to Fail problem}

Several recent papers have argued that the Milky Way classical dwarfs, in the context of $\Lambda$CDM, must inhabit the most massive DM subhalos before infall \citep[e.g.][]{2018MNRAS.473.2060J,2017arXiv171106267K,2018arXiv180707093R}. However, these massive subhalos have central densities that are too high to be consistent with the observed stellar velocity dispersions of the Milky Way classical dwarfs \citep[e.g.][]{2006MNRAS.367..387R}, a problem that has become known as `Too Big to Fail' \citep{2011MNRAS.415L..40B}. 

The nomenclature `Too Big to Fail' (hereafter TBTF) refers to the fact that TBTF is solved if the most massive subhalos are devoid of stars and gas, placing the classical dwarfs instead in lower mass and, therefore, lower density subhalos. However, such a solution is puzzling because it requires the most massive subhalos to end up dark while their lighter cousins form stars. Such massive subhalos ought to be `Too Big to Fail'.

An alternative solution to TBTF is that the central density of the most massive subhalos is lower than expected from pure DM structure formation simulations in $\Lambda$CDM \citep{2006MNRAS.367..387R}. Indeed, \citet{2012MNRAS.422.1203B} point out that TBTF can be cast as a `central density problem', akin to the cusp-core problem for isolated dwarfs (see \S\ref{sec:intro}).

With the results of this paper, we are now in a position to revisit TBTF. \citet{2011MNRAS.415L..40B} argue that, statistically, $2-4$ of the Milky Way classical dwarfs have an unexpectedly low central density. From Figure \ref{fig:dmheating}, middle panel, of the satellite dwarfs studied here, only Fornax has a central density that is lower than expected in pure DM structure formation simulations in $\Lambda$CDM (compare the location of Fornax with the grey band on this plot). However, the Sagittarius dSph also appears to inhabit a massive pre-infall subhalo \citep{2017MNRAS.464..794G,2018arXiv180707093R}. If Fornax and Sagittarius inhabit massive pre-infall halos (with $M_{200} > 10^{10}$\,M$_\odot$), then this is already sufficient to significantly alleviate the Milky Way's TBTF problem. However, in addition to Fornax and Sagittarius, there may have been other Fornax-like galaxies that fell in late and did not survive. As discussed in \citet{2016MNRAS.459.2573R}, early infalling dwarfs have their star formation shut down before they can fully transform their cusp to a core. Indeed, as we have shown in this paper, the Milky Way dSphs with only old-age stars are consistent with this (see Figure \ref{fig:dmheating}, middle panel, black data points). By contrast, late infalling dwarfs have time to transform their cusps to cores, becoming more susceptible to tidal destruction than expected in pure DM structure formation simulations. A full solution to TBTF may require some of these late infalling cored dwarfs to be tidally destroyed \citep[e.g.][]{2012ApJ...761...71Z,2014ApJ...786...87B,2016ApJ...827L..23W}. We will study this in more detail in future work.

\subsection{The nature of DM}\label{sec:dmnature}

Our \GravSphere\ models favour a wide range of central DM densities in dwarfs with similar $M_*$ (Figure \ref{fig:dmheating}, left panel). Furthermore, the densest dwarfs are those whose star formation shut down long ago, while the low density dwarfs have more extended star formation. These results are in excellent agreement with models in which cold DM `heats up' at the centres of dwarf galaxies due to bursty star formation (Figure \ref{fig:dmheating}, middle and right panels). However, they are challenging to understand in models where large DM cores are ubiquitous. Many modifications to the nature of DM have been proposed to explain the observed DM cores in dwarf irregular galaxies (see \S\ref{sec:intro}). However, these typically produce DM cores in {\it all} dwarfs, which is not what we find here. In a companion paper, we used our densest dwarf, Draco, to place a new constraint on the DM self-interaction cross section \citep{Read:2018pft}; dense dwarfs like Draco can now be used to place similar constraints on any model that produces ubiquitous DM cores (e.g. ultra-light axion DM; \citealt{2015MNRAS.451.2479M,2017MNRAS.472.1346G}).

\section{Conclusions}\label{sec:conclusions}

%MOREMORE CHECK THIS SECTION FOR CLARITY WRT SFH => HEATING 

We have used stellar kinematics and HI rotation curves to infer the radial DM density profile of eight dwarf spheroidal (dSph) and eight dwarf irregular (dIrr) galaxies with a wide range of star formation histories. Our key findings are as follows:

\begin{itemize}

\item The dwarfs fell into two distinct classes. Galaxies with only old stars ($>6$\,Gyrs old) had central DM densities, $\rho_{\rm DM}(150\,{\rm pc}) > 10^8$\,M$_\odot$\,kpc$^{-3}$, consistent with DM cusps; those with star formation until at least 3\,Gyrs ago had $\rho_{\rm DM}(150\,{\rm pc}) < 10^8$\,M$_\odot$\,kpc$^{-3}$, consistent with DM cores (Figure \ref{fig:dmheating}, left panel).

\item We estimated pre-infall halo masses for our sample of dwarfs, using HI rotation curve measurements for the dIrr sample and abundance matching for the dSph sample. With this, we showed that their $\rho_{\rm DM}(150\,{\rm pc})$ as a function of $M_{200}$ is in good agreement with models in which DM is kinematically `heated up' by bursty star formation. The dwarfs with only old-age stars lay along the track predicted by the NFW profile in $\Lambda$CDM, consistent with having undergone no measurable DM heating. By contrast, those with extended star formation lay along the track predicted by the \coreNFW\ profile from \citet{2016MNRAS.459.2573R}, consistent with maximal DM heating (Figure \ref{fig:dmheating}, middle panel).

\item We found that $\rho_{\rm DM}(150\,{\rm pc})$ for our sample of dwarfs is anti-correlated with their stellar mass to pre-infall halo mass ratio, $M_*/M_{200}$ (Figure \ref{fig:dmheating}, right panel). This is also in good quantitative agreement with predictions from recent DM heating models \citep{2012ApJ...759L..42P,2014MNRAS.437..415D,2015MNRAS.454.2981C,2016MNRAS.459.2573R,2016MNRAS.456.3542T}.

\item While not the main focus on this paper, in Appendix \ref{app:dwarftwins} we discussed the implications of our results for `alternative gravity' models for DM. There, we showed that the dwarf `twins' Draco and Carina provide a particularly clean test of such models. These two dwarfs have similar $M_*$, $R_{1/2}$ and orbit around the Milky Way, yet favour very different dark matter density profiles. In $\Lambda$CDM, this is explained by Carina and Draco inhabiting halos with different pre-infall masses and concentrations (Figure \ref{fig:dmheating}, middle panel). In alternative gravity theories, however, the existence of visibly similar galaxies with different gravitational force-fields represents a major challenge (Figure \ref{fig:mond}).

\end{itemize}

\section{Acknowledgments}
We would like to thank Arianna Di Cintio, Stacy McGaugh, Simon White and Hosein Haghi for useful feedback on an early draft of this paper. We would like to thank the referee, Andrew Pontzen, for a constructive report that improved the clarity of the paper. We would also like to thank N. McMonigal for kindly providing the photometric data for Carina from \citet{mcmonigal14}. JIR would like to thank the KITP in Santa Barbara and the organisers of the ``The Small-Scale Structure of Cold(?) Dark Matter'' programme. This paper benefited from helpful discussions that were had during that meeting. JIR would like to acknowledge support from SNF grant PP00P2\_128540/1, STFC consolidated grant ST/M000990/1 and the MERAC foundation. This research was supported in part by the National Science Foundation under Grant No. NSF PHY-1748958. M.G.W. is supported by National Science Foundation grant AST-1412999.

The Pan-STARRS1 Surveys (PS1) and the PS1 public science archive have been made possible through contributions by the Institute for Astronomy, the University of Hawaii, the Pan-STARRS Project Office, the Max-Planck Society and its participating institutes, the Max Planck Institute for Astronomy, Heidelberg and the Max Planck Institute for Extraterrestrial Physics, Garching, The Johns Hopkins University, Durham University, the University of Edinburgh, the Queen's University Belfast, the Harvard-Smithsonian Center for Astrophysics, the Las Cumbres Observatory Global Telescope Network Incorporated, the National Central University of Taiwan, the Space Telescope Science Institute, the National Aeronautics and Space Administration under Grant No. NNX08AR22G issued through the Planetary Science Division of the NASA Science Mission Directorate, the National Science Foundation Grant No. AST-1238877, the University of Maryland, Eotvos Lorand University (ELTE), the Los Alamos National Laboratory, and the Gordon and Betty Moore Foundation.

\appendix

\section{DM density profiles of the dwarf irregular galaxies}\label{app:dIrrs}

\begin{figure*}
\begin{center}
\includegraphics[width=0.75\textwidth]{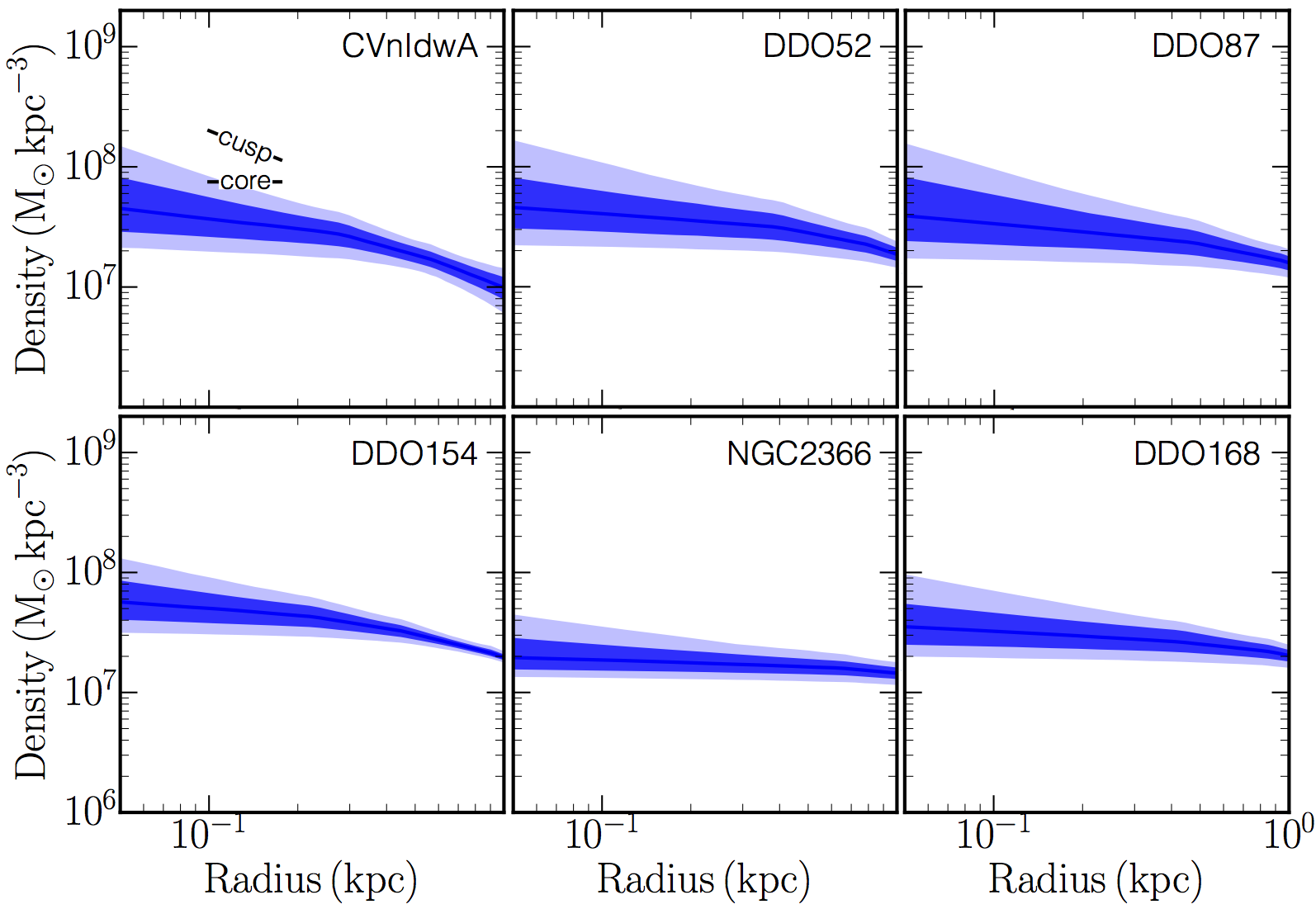}
\caption{As Figure \ref{fig:killer_plot}, but for the remaining dIrrs. These galaxies have all actively been forming stars over the past 0.1\,Gyrs \citep{2012AJ....143...47Z}, but do not have star formation histories measured from deep colour magnitude diagrams. For this reason, we show just their radial dark matter density profiles.}
\label{fig:killer_plot_dIrrs}
\end{center}
\end{figure*}

In this Appendix, we show the DM density profile results for the remainder of our dIrr sample (Figure \ref{fig:killer_plot_dIrrs}). These dIrrs are all actively forming stars today \citep{2012AJ....143...47Z}, but do not have star formation histories measured from deep colour magnitude diagrams (see \S\ref{sec:data}). Notice that all of them are consistent with having constant density dark matter cores inside $\sim 500$\,pc. Even those that permit steeper profiles within their 95\% confidence intervals (e.g. CVnIdwA and DDO87) have central densities that are systematically lower than all of the dSphs, except Fornax (see Figures \ref{fig:killer_plot}, \ref{fig:killer_plot_umisex} and \ref{fig:dmheating}).

\section{Dwarf twins: a challenge for alternative gravity theories}\label{app:dwarftwins}

\begin{figure}
\begin{center}
\includegraphics[width=0.45\textwidth]{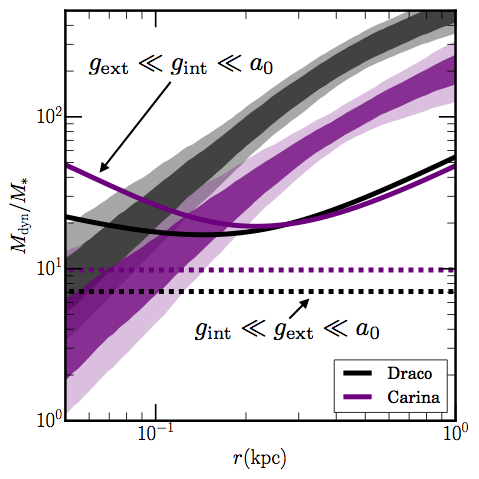}
\caption{The dwarf `twins' Carina and Draco: a challenge for alternative gravity explanations for DM. The contours show the 68\% (dark) and 95\% (light) confidence intervals of the ratio of the dynamical to the stellar mass, $M_{\rm dyn}/M_*$, for Draco (black) and Carina (purple), calculated from our \GravSphere\ model chains. The solid and dashed black and purple lines show predictions for Draco and Carina in MOND in two limiting `deep MOND' regimes, as marked (equations \ref{eqn:mond_general} and \ref{eqn:mond_ext}). In all cases, the MOND predictions show poor agreement with our dynamical inferences. More troublesome, however, is the similarity of the predictions for both galaxies. Their $M_*$, $R_{1/2}$ and distance from the Milky Way lead to similar predictions for $M_{\rm dyn}/M_*$ in MOND. Yet, their stellar kinematics imply that Draco is substantially denser than Carina. This is challenging to understand in any alternative gravity theory that seeks to fully explain DM, not just MOND.}
\label{fig:mond}
\end{center}
\end{figure}

While not the main focus of this paper, in this Appendix we consider what our results imply for `alternative gravity' theories of DM. In these theories, weak-field gravity is altered at low acceleration or on large scales to explain flat rotation curves and anomalous galactic dynamics without invoking an invisible dark matter \citep[e.g.][]{1983ApJ...270..365M,2004PhRvD..70h3509B,2016arXiv161102269V}.

Many of the original alternative gravity theories like MOdified Newtonian Dynamics (MOND; \citealt{1983ApJ...270..365M}) and TeVeS \citep{2004PhRvD..70h3509B} have now been ruled out as a complete explanation for DM by data from the cosmic microwave background radiation and large scale structure \citep[e.g.][]{2006PhRvL..96a1301S,2011IJMPD..20.2749D}, and galaxy clusters \citep[e.g.][]{2006ApJ...648L.109C,2008MNRAS.389..250N}. However, modern versions of these theories revert to a $\Lambda$CDM-like cosmology on large scales, thereby sidestepping these constraints \citep[e.g.][]{2009PhRvD..80f4007L,2015PhRvD..91b4022K,2016PhRvD..93j3533K}. This makes it interesting to test modifications to Newtonian gravity in the weak-field regime where alternative gravity theories have traditionally had more success \citep[e.g.][]{2012LRR....15...10F,2017ApJ...836..152L}. In this Appendix, we show that the `dwarf twins' Carina and Draco offer us a particularly clean test of such modified weak-field gravity theories.

The idea of using pairs of similar dwarfs to test modified gravity theories was first suggested by \citet{2013ApJ...775..139M}. They compared dwarfs with similar stellar mass and external tidal field orbiting around M31, finding that the pairs they considered were consistent with predictions in MOND. However, the orbits of the M31 dwarfs are not known, allowing some leeway in explaining pairs that do not precisely match up. By contrast, the Milky Way dwarfs Draco and Carina present a particularly clean test because of their similar stellar masses, half stellar mass radii, distances from the Milky Way\footnote{Note that UMi and Sextans could also be good `twin' candidates for Draco, however the uncertainties on their dynamical masses are larger than for Carina due to their smaller number of radial velocity measurements. Aquarius is also a promising `twin' for Leo II, but taking into account its gas mass, its baryonic mass is actually substantially larger than Leo II's (see Table \ref{tab:data}). Aquarius also orbits in a much weaker tidal field and may be flattened by rotation \citep[e.g.][]{2016MNRAS.459.2573R}. For these reasons, of the dwarfs we study here, Draco and Carina are the cleanest `twins' for testing alternative gravity models.} (Table \ref{tab:data}), and orbits \citep{2010MNRAS.406.2312L,2018arXiv180409381G}.

We now show quantitatively that Draco and Carina do indeed present a challenge for alternative gravity theories, using MOND as an example. Assuming spherical symmetry, the MOND force field, ${\bf g}$, relates to the standard Newtonian force field, ${\bf g}_N$, as \citep[e.g.][]{2012LRR....15...10F}:

\begin{equation} 
{\bf g} = {\bf g}_N\frac{\left(1+\sqrt{1+\frac{4a_0^2}{|{\bf g}_N|^2}}\right)^{1/2}}{\sqrt{2}}
\label{eqn:mond_general}
\end{equation}
where $a_0 \sim 1.2 \times 10^{10}$\,m\,s$^{-2}$ is the MOND acceleration scale.

Unlike Newtonian gravity, MOND is not a linear theory and so we must worry about how the force field from the Milky Way influences the dynamics of stars moving in Draco and Carina \citep[e.g.][]{2012LRR....15...10F,2014MNRAS.440..746A}. This is called the `external field effect'. Fortunately, these two galaxies are to a very good approximation in the `deep MOND' regime. Using the recent Milky Way model from \citet{2017MNRAS.465...76M}\footnote{We calculate the enclosed mass as a function of radius for this model using the \url{https://github.com/PaulMcMillan-Astro/GalPot} code.}, the magnitude of the acceleration from the Milky Way at 100\,kpc is $g_{\rm ext} \sim 10^{-11}$\,m\,s$^{-2}$ which is a factor of ten smaller than $a_0$. Similarly, the internal acceleration at 150\,pc for Draco is $g_{\rm int} \sim 7 \times 10^{-12}$\,m\,s$^{-2}$. The dynamics in this deep MOND limit then fall into two limiting cases: the `quasi-Newtonian' regime, where $g_{\rm int} \ll g_{\rm ext} \ll a_0$; and the isolated regime, where $g_{\rm ext} \ll g_{\rm int} \ll a_0$ \citep[e.g.][]{2014ApJ...785..166D}. Carina and Draco lie closer to the quasi-Newtonian regime than the isolated regime, but we will calculate results for both to show these two extremum cases.

In the quasi-Newtonian regime, the dynamics become Newtonian but with a modified gravitational constant, $G \rightarrow G g_{\rm ext} / a_0$ \citep{2014ApJ...785..166D}. In this case, the ratio of the dynamical mass to the stellar mass becomes:

\begin{equation}
\frac{M_{\rm dyn}}{M_*} = \frac{g_{\rm ext}}{a_0} = {\rm const.}
\label{eqn:mond_ext}
\end{equation}
where $g_{\rm ext}$ will be slightly different for Draco and Carina due to their different distances from the Milky Way centre (see Table \ref{tab:data}).

In the isolated regime, $|{\bf g}_N| \ll a_0$ and from equation \ref{eqn:mond_general} we obtain:

\begin{equation}
\frac{M_{\rm dyn}}{M_*} \simeq \sqrt{\frac{a_0}{G M_*(r)}} r
\label{eqn:mond}
\end{equation}
Using the best-fit $M_*(r)$ from the \GravSphere\ model fits to the projected light profiles of Draco and Carina, in Figure \ref{fig:mond} we show predictions for $M_{\rm dyn}/M_*$ for Draco and Carina in MOND. We show results for both the isolated regime (solid lines) and the quasi-Newtonian regime (dashed lines), as marked on the Figure. The contours show the 68\% (dark) and 95\% (light) confidence intervals of the ratio of the dynamical to the stellar mass, $M_{\rm dyn}/M_*$, for Draco (black) and Carina (purple) calculated from our \GravSphere\ model chains. Notice that in all cases, the MOND predictions show poor agreement with our dynamical inferences. Indeed, it has been noted in the literature before that Draco \citep{1992ApJ...397...38G,2001ApJ...563L.115K,2007ApJ...667..878S,2010ApJ...722..248M,2017ApJ...835..233A} and Carina \citep{2008MNRAS.387.1481A,2017ApJ...835..233A} are poorly fit by MOND, even when accounting for the external field effect and tides \citep{2014MNRAS.440..746A}. Here, we point out an even more severe problem: these two galaxies require {\it different dynamical mass profiles for almost the same radial light profile}. This is a challenge not only for MOND, but for {\it any} weak-field gravity theory that seeks to fully explain DM. 

In the context of $\Lambda$CDM, the different central densities of Carina and Draco can be understood as owing to their different pre-infall halo masses and concentrations (see Figure \ref{fig:dmheating}). In alternative gravity theories, the only possible explanation is that either one or both of these galaxies is not in dynamical equilibrium. However, at least in MOND, such an explanation is problematic. \citet{2000ApJ...541..556B} showed that Draco and Carina will be largely immune to tidal effects in MOND if their orbital pericentres are $r_p \simgt 32$\,kpc and $r_p \simgt 41$\,kpc, respectively. The latest proper motion data from {\it Gaia} DR2 for these two galaxies (assuming the Milky Way `model 2' from \citealt{2018arXiv180409381G}) gives $r_p = 32^{+6.1}_{-5.3}$\,kpc for Draco and $r_p = 74.5^{+23.7}_{-19.5}$\,kpc for Carina, suggesting only a weak tidal influence from the Milky Way. Indeed, observational evidence for tides -- in the form of a velocity gradient, inflated velocity dispersion at large radii and/or feature in the photometric light profile -- has been reported only for Carina \citep{2006ApJ...649..201M} and even this is contested \citep{mcmonigal14}. Finally, \citet{2014MNRAS.440..746A} presented a detailed numerical calculation of the effect of tides on satellite galaxies in MOND. They found that tides are unable to sufficiently inflate the velocity dispersion of Carina in MOND to explain the data. While they did not explicitly model Draco, they showed that lowering the pericentre for their Carina models led to more tidal stripping, lowering the mass of Carina and, ultimately, {\it lowering} its velocity dispersion. It seems that, at least in MOND, it is not possible to simultaneously explain the data for Carina and Draco. 

\section{Varying the scale at which we estimate the inner DM density}\label{app:scale_test}

\begin{figure*}
\begin{center}
\includegraphics[width=0.95\textwidth]{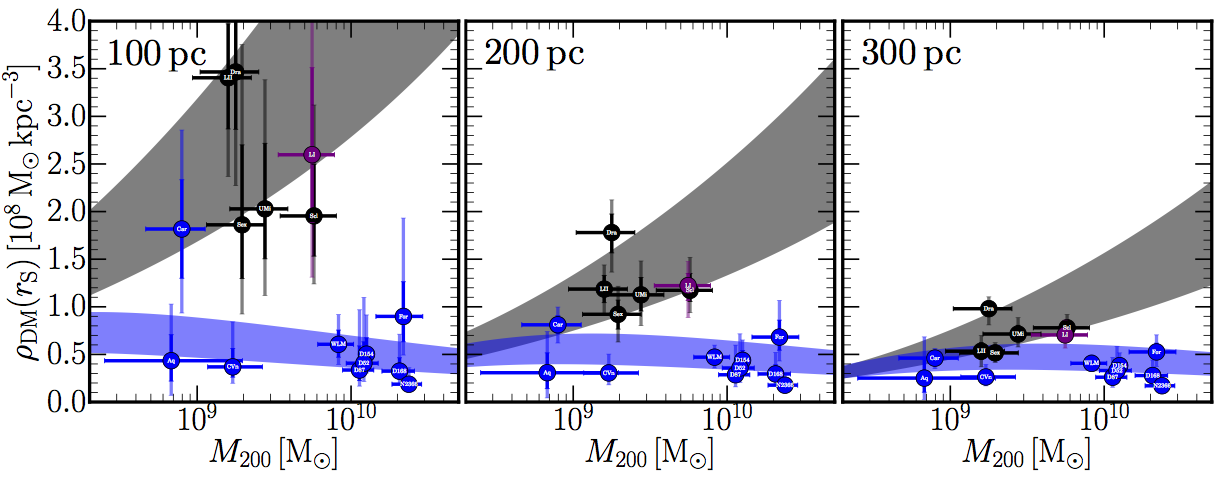}
\vspace{-1mm}
\caption{Varying the scale at which we estimate the inner DM density. From left to right, the panels show the inner DM density, $\rho_{\rm DM}(r_{\rm S})$, for $r_{\rm S} = 100, 200$ and $300$\,pc, as marked. The data points and contours are as in Figure \ref{fig:dmheating}.}
\label{fig:robust}
\end{center}
\end{figure*}

In this Appendix, we show how our results change if we vary the scale at which we estimate the inner DM density. In Figure \ref{fig:robust}, we show the inner DM density, $\rho_{\rm DM}(r_{\rm S})$, for $r_{\rm S} = 100, 200$ and $300$\,pc, as marked on the panels. The data points and contours are as in Figure \ref{fig:dmheating}. As can be seen, our results are not altered by the choice of $r_{\rm S}$. For $r_{\rm S} = 100$\,pc (left panel), we still see a clear separation in density between those dwarfs that stopped forming stars long ago (black) and those that formed stars until recently (blue). However, the uncertainties on $\rho_{\rm DM}(100\,{\rm pc})$ are larger than for our default choice of $\rho_{\rm DM}(150\,{\rm pc})$. As $r_{\rm S}$ is increased, the error bars on $\rho_{\rm DM}(r_{\rm S})$ shrink, but so too does the difference between cusped and cored models in this space. Our default choice of $r_{\rm S} = 150\,{\rm pc}$ represents a compromise between minimising the error on $\rho_{\rm DM}(r_{\rm S})$ and maximising the difference between cusped and cored models.

\section{\GravSpherebf\ constraints on the logarithmic slope of the inner DM density profile}\label{app:gammaDM}

In this Appendix, we present our \GravSphere\ model inference of $\gamma_{\rm DM}(150\,{\rm pc})$ for our sample of dwarfs. Recall that in \citet{Read:2018pft}, we showed that $\gamma_{\rm DM}(150\,{\rm pc})$ depended on our choice of priors on $\gamma_{\rm DM}$. To show this, we introduced a rather extreme prior on $\gamma_{\rm DM}$ designed to explicitly bias our models towards cores. We assumed a flat prior over the range $-3 < \gamma_{\rm DM}' < 2$, setting $\gamma_{\rm DM} = 0$ if $\gamma_{\rm DM}' > 0$ and $\gamma_{\rm DM} = \gamma_{\rm DM}'$ otherwise. In the absence of constraining data, this `AltGam' prior biases \GravSphere\ towards cores by creating a large region of hypervolume in which $\gamma_{\rm DM} = 0$. (Note that we consider this prior to be extreme, using it only to test our sensitivity to priors on $\gamma_{\rm DM}$.) 

\begin{figure}
\begin{center}
\includegraphics[width=0.47\textwidth]{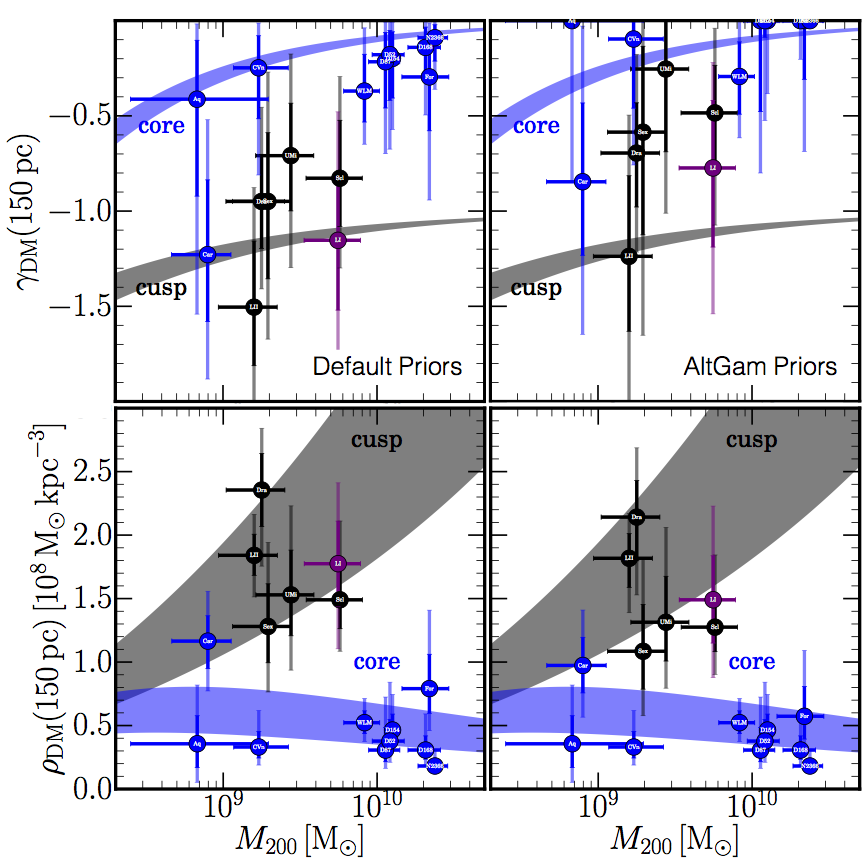}
\vspace{-1mm}
\caption{The central logarithmic cusp slope of the dark matter density profile -- $\gamma_{\rm DM}(150\,{\rm pc}) \equiv d\ln\rho_{\rm DM}/d\ln r(150\,{\rm pc})$ -- as a function of $M_{200}$ using our default priors on $\gamma_{\rm DM}$ (top left) and using an extreme prior designed to bias our \GravSphere\ models towards cores (`AltGam'; top right panel and see text for details). The colour of the points is as in Figure \ref{fig:dmheating}. The grey and blue bands bracket the extremum cases of no cusp-core transformation and complete cusp-core transformation in $\Lambda$CDM, respectively (c.f. the similar bands in Figure \ref{fig:dmheating}, middle panel). The bottom two panels show similar results for $\rho_{\rm DM}(150\,{\rm pc})$ using our default priors (left) and the AltGam priors (right). Notice that \GravSphere's inference of $\gamma_{\rm DM}(150\,{\rm pc})$ is affected by the priors on $\gamma_{\rm DM}$, while its inference of $\rho_{\rm DM}(150\,{\rm pc})$ is not. However, the ordering of $\gamma_{\rm DM}(150\,{\rm pc})$ is unaffected by the priors: dwarfs that have only old-age stars (black data points) are systematically steeper than those with a younger stellar population (blue data points).}
\label{fig:altgam}
\end{center}
\end{figure}

In Figure \ref{fig:altgam}, we show our inference of $\gamma_{\rm DM}(150\,{\rm pc})$ for our default priors on $\gamma_{\rm DM}$ (left) and using the above AltGam prior (right). The bottom panels show the corresponding results for $\rho_{\rm DM}(150\,{\rm pc})$. Similarly to our findings in \citet{Read:2018pft}, our results for $\gamma_{\rm DM}(150\,{\rm pc})$ depend on our priors, whereas $\rho_{\rm DM}(150\,{\rm pc})$ is more robust. This is why we focus throughout this paper on our inference of $\rho_{\rm DM}(150\,{\rm pc})$ rather than $\gamma_{\rm DM}(150\,{\rm pc})$. Nonetheless, while $\gamma_{\rm DM}(150\,{\rm pc})$ systematically shifts with our prior, the ordering of the dwarfs remains unchanged. Notice that dwarfs with old-age stars (black data points) are systematically steeper at 150\,pc than those with younger stellar populations (blue data points). This is consistent with our findings for $\rho_{\rm DM}(150\,{\rm pc})$.

\bibliographystyle{mnras}
\bibliography{final_refs}

\end{document}